\documentclass[useAMS,usenatbib]{mnras}
\usepackage{graphicx,epsfig,longtable}
\usepackage{epstopdf}
\usepackage{xcolor}
\usepackage{amsmath,amssymb}
\usepackage{tabularx} 

\usepackage[T1]{fontenc}

\usepackage[utf8]{inputenc}

\usepackage[]{natbib}
\usepackage[utf8]{inputenc}
\usepackage{booktabs}
\let\textbf\relax

\newcommand{\Msun}{$M_{\sun}$}
\newcommand{\vecx}{\mathbf{x}}
\newcommand{\vecn}{\mathbf{n}}
\newcommand{\vecF}{\mathbf{F}}
\newcommand{\vecv}{\mathbf{v}}
\newcommand{\tenP}{\mathbf{P}}
\newcommand{\XLA}{X_{\mathrm{LA}}}
\newcommand{\dd}{\mathrm{d}}
\newcommand{\fth}{f_{\mathrm{th}}}

\newcommand{\Qheat}{Q_{\mathrm{heat}}}
\newcommand{\Qneu}{Q_{\mathrm{neu}}}
\newcommand{\kapla}{\kappa_{\rm LA}}
\newcommand{\thyd}{t_{\rm hyd}}
\newcommand{\tkn}{t_{\rm KN}}
\newcommand{\Mej}{M_{\rm ej}}
\newcommand{\theobs}{\theta_{\rm obs}}

\title[Dynamical ejecta with weak processes II: Kilonova]{Dynamical ejecta of neutron star mergers with nucleonic weak processes II: Kilonova emission}
\author[O.~Just et al.]
  {O.~Just$^{1,2}$, I. Kullmann$^3$, S.~Goriely$^3$, A. Bauswein$^1$, H.-T. Janka$^4$, \newauthor C.~E.~Collins$^1$ \\ 
  $^1$GSI Helmholtzzentrum f\"ur Schwerionenforschung, Planckstrasse 1, 64291 Darmstadt, Germany\\
  $^2$Astrophysical Big Bang Laboratory, RIKEN Cluster for Pioneering Research, 2-1 Hirosawa, Wako, Saitama 351-0198, Japan \\
  $^3$Institut d'Astronomie et d'Astrophysique, CP-226, Universit\'e Libre de Bruxelles, 1050 Brussels, Belgium\\
  $^4$Max-Planck-Institut f\"ur Astrophysik, Postfach 1317, 85741 Garching, Germany 
 }
\date{Released 2021 Xxxxx XX}

\def\ga{\,\,\raise0.14em\hbox{$>$}\kern-0.76em\lower0.28em\hbox
{$\sim$}\,\,}
\def\la{\,\,\raise0.14em\hbox{$<$}\kern-0.76em\lower0.28em\hbox
{$\sim$}\,\,}
\def\Msun{$M_{\odot}$}

\begin{document}

\label{firstpage}

\maketitle

\begin{abstract}
  The majority of existing results for the kilonova (or macronova) emission from material ejected during a neutron-star (NS) merger is based on (quasi-)one-zone models or manually constructed toy-model ejecta configurations. In this study we present a kilonova analysis of the material ejected during the first $\sim 10\,$ms of a NS merger, called dynamical ejecta, using directly the outflow trajectories from general relativistic smoothed-particle hydrodynamics simulations including a sophisticated neutrino treatment and the corresponding nucleosynthesis results, which have been presented in Part I of this study. We employ a multi-dimensional two-moment radiation transport scheme with approximate M1 closure to evolve the photon field and use a heuristic prescription for the opacities found by calibration with atomic-physics based reference results. We find that the photosphere is generically ellipsoidal but augmented with small-scale structure and produces emission that is about 1.5-3 times stronger towards the pole than the equator. The kilonova typically peaks after $0.7-1.5\,$days in the near-infrared frequency regime with luminosities between $3-7\times 10^{40}\,$erg\,s$^{-1}$ and at photospheric temperatures of $2.2-2.8\times 10^3\,$K. A softer equation of state or higher binary-mass asymmetry leads to a longer and brighter signal. Significant variations of the light curve are also obtained for models with artificially modified electron fractions, emphasizing the importance of a reliable neutrino-transport modeling. None of the models investigated here, which only consider dynamical ejecta, produces a transient as bright as AT2017gfo. The near-infrared peak of our models is incompatible with the early blue component of AT2017gfo.
\end{abstract}


\section{Introduction}

The recent gravitational-wave observation of a binary neutron-star (NS) merger, GW170817 \citep[e.g.][]{Abbott2017q}, together with its optical and near-infrared (IR) electromagnetic counterparts, AT2017gfo \citep[e.g.][]{Kasen2017a, Nicholl2017m, Smartt2017s, Tanaka2017t, Tanvir2017a, Villar2017a}, provides long-sought observational support to the idea \citep{Lattimer1976, Eichler1989} that a substantial fraction of material expelled during NS mergers undergoes the rapid neutron-capture (or r-) process \citep[see, e.g.,][for recent reviews]{Frebel2018m, Kajino2019b, Arnould2020f, Cowan2021g}, which is responsible for the nucleosynthesis of about half of the trans-iron elements in the Universe. Major evidence for this interpretation, hence for the interpretation of AT2017gfo as a so-called kilo- or macronova \citep[e.g.][]{Li1998, Kulkarni2005, Metzger2010c, Metzger2019a}, comes from the fact that the light curve of AT2017gfo declined at early times, $t$, roughly as $t^{-1.3}$ characteristic of the energy release rate connected to the consecutive radioactive decay of newly forged r-process elements \citep{Metzger2010c, Roberts2011, Goriely2011}. Additional support is provided by spectroscopic features of AT2017gfo, which point to the existence of strontium \citep{Watson2019s} and lanthanides in the ejecta \citep{Kasen2017a, Tanaka2017t}. Moreover, the appearance of (at least) two different, red and blue, components in AT2017gfo suggests the existence of multiple channels of matter ejection. Yet, no uniform consensus is reached as to what ejecta component was (mainly) responsible for which kilonova component. The most dominant types of ejecta are the dynamical ejecta \citep[e.g.][]{Ruffert1996a, Rosswog1999, Goriely2011, Korobkin2012, Bauswein2013, Sekiguchi2015a, Palenzuela2015a, Foucart2016a, Radice2018b}, neutrino-driven winds \citep[e.g.][]{Metzger2014, Perego2014a, Just2015a, Fujibayashi2018a, Foucart2020a}, and ejecta driven by turbulent viscosity \citep[e.g.][]{Fernandez2013b, Just2015b, Siegel2017b, Fujibayashi2018a, Miller2019a}.

The light curve of a kilonova delivers unique information about the outflow mass, velocity, and composition to the observer. These properties are of paramount importance for, among others, chemical evolution models that attempt to unravel the dominant sites of heavy-element nucleosynthesis \citep[e.g.][]{Shen2015, Vangioni2016, Cote2018i, van-de-Voort2020r}, in order to understand the evolution and lifetime of the NS remnant and the nuclear equation of state (EoS) \citep[e.g.][]{Bauswein2013, Hotokezaka2013b, Lippuner2017a, Bauswein2017b, Margalit2017a, Radice2018e, Rezzolla2018a, Coughlin2018a, Shibata2019b}, or to constrain the processes that launch and collimate ultrarelativistic jets of short gamma-ray bursts \citep[e.g.][]{Just2016, Ito2021a, Hamidani2020r, Gottlieb2021o}. However, deciphering an observed kilonova light curve remains an ambitious endeavor because of the complexity of involved physics ingredients. Profound theoretical understanding is required of the neutrino-magneto-hydrodynamical matter ejection processes that shape the ejecta structure and set the neutron density (cf. aforementioned references for different ejecta components), of nuclear reaction networks that determine the elemental abundances and the radioactive heating rates powering the emission \citep[e.g.][]{Goriely2013, Mumpower2015, Lippuner2017b, Barnes2021c}, and of atomic physics and radiative processes that define the opacities and observer-angle dependent emission rates in various frequency bands \citep[e.g.][]{Kasen2013, Tanaka2020a, Fontes2020d}.

While tremendous progress has been made in the recent years, each of these aspects still faces considerable challenges and uncertainties. As for the dynamical ejecta, a long-standing uncertainty is connected to their lanthanide fraction and the question whether they would produce rather a red or a blue kilonova, or could even produce both. The lanthanide fraction resulting after the r-process is determined mainly by the electron fraction, or proton-to-baryon ratio, which, in turn, is set during the matter ejection process as a result of the competing emission and absorption processes of electron neutrinos and antineutrinos. Early NS merger simulations, which neglected neutrinos \citep{Bauswein2013} or included neutrino emission but ignored absorption \citep{Freiburghaus1999, Korobkin2012}, found a solar-like abundance pattern for all elements with mass numbers above $A\sim 130-140$ and thus a high lanthanide fraction and correspondingly high opacity. These results implied a kilonova that peaks on rather long timescales and in red spectral colors \citep{Tanaka2013, Grossman2014}. Subsequent models using various approximate prescriptions for neutrino absorption revealed that the composition is not as robust as previously thought and may sensitively depend on neutrino-transport effects \citep{Wanajo2014a, Goriely2015, Martin2018a, Radice2018b}. However, since models including neutrino absorption are still rare and all of them still employ some kind of approximations, and since self-consistent kilonova calculations based on these neutrino-hydrodynamics models are even rarer, the question of the kilonova signature of the dynamical ejecta remains open.

In this study, we take a step towards resolving this question by investigating the kilonova that results from dynamical ejecta modeled by using a state-of-the-art neutrino scheme, namely the Improved Leakage-Equilibration-Absorption Scheme (ILEAS; \citealp{Ardevol-Pulpillo2019a}), which in particular includes neutrino-absorption effects. The present study follows up on the nucleosynthesis analysis presented in \citet{Kullmann2021x} (called ``Part I'' hereafter) and discusses the kilonova emission for exactly the same models.

Many studies, in fact most of the discovery studies connected to GW170817, are based on a simple but powerful one-zone approximation \citep[e.g.][]{Arnett1982}, which models the ejecta as an expanding homogeneous sphere with a constant, grey (i.e. frequency-independent) opacity. This approach, or similarly simple considerations \citep[e.g.][]{Grossman2014, Metzger2019a, Hotokezaka2020a}, are often employed in a post-processing step to estimate the kilonova emission based on the bulk properties (total mass, average velocity, average electron fraction) of ejecta obtained in hydrodynamical simulations. Alternatively, schemes are employed that solve the radiative-transfer equations very accurately using state-of-the-art opacity descriptions \citep[][]{Kasen2013, Tanaka2013, Wollaeger2018a, Bulla2019a}, but which often assume manually constructed toy-model distributions of the ejecta and that are numerically involved and computationally expensive. In this paper we apply a scheme, based on the M1 approximation of radiative transfer, that in terms of accuracy and complexity fills the gap between the two aforementioned approaches. Due to its close relationship to hydrodynamics, it can be implemented more readily in existing hydrodynamics codes than schemes solving the full Boltzmann equation. As opposed to most existing studies, we adopt the outflow distributions directly from neutrino-hydrodynamics simulations, and we take the composition and the radioactive heating rate from the consistently post-processed nucleosynthesis tracers.

This paper is organized as follows: In Sect.~\ref{sec:methodology} we briefly review the hydrodynamic models and corresponding nucleosynthesis results from Part I, describe the employed mapping from hydrodynamic tracers to the velocity space, and we outline the governing equations and additional assumptions that enter the kilonova evolution scheme. In Sect.~\ref{sec:results}, we present and compare the resulting kilonova light curves. Subsequently, in Sect.~\ref{sec:discussion} we briefly speculate about the implications of our models to the interpretation of AT2017gfo, point out the advantages of a tracer-based scheme compared to one-zone models, and contrast our results with some published works. Finally, we summarize in Sect.~\ref{sec:summary-conclusions}.

\section{Methodology}\label{sec:methodology}

\newlength{\mytabcolsep}
\setlength{\mytabcolsep}{\tabcolsep}
\setlength{\tabcolsep}{4pt}
\begin{table*}
\centering
\caption{Global properties of our models as measured in the entire volume (total), measured only in the polar regions defined by the cones of opening angle $\pi/4$ around the two poles (polar), and measured in the remaining equatorial sub-volume (equat.), namely the ejecta mass, average velocity, average electron fraction, average lanthanide mass fraction, average initial opacity, as well as the time, bolometric luminosity, surface-averaged photospheric temperature, and surface-averaged photospheric velocity of the resulting kilonova at peak epoch. We define the peak time as the first time when the total luminosity equals the total heating rate, and all other peak quantities are measured at that time. Note that a spherically symmetric mass distribution would correspond to $M_{\rm ej}^{\rm polar}/M_{\rm ej}^{\rm total}=0.292$.}
\begin{tabular}{llccccccccc}
\hline \hline
model               & region & $M_{\rm ej}$                       & $\langle v/c\rangle$ & $\langle Y_{e,\rho_{\rm net}}\rangle$ & $\langle X_{\rm LA}\rangle$ & $\langle\kappa_{\rm LA}\rangle$ & $t_{\rm peak}$ & $L_{\rm peak}$           & $\langle T_{\rm ph} \rangle_{\rm peak}$ & $\langle v_{\rm ph}/c \rangle_{\rm peak}$ \\
                    &        & [\Msun ($M_{\rm ej}^{\rm total}$)] &                      &                                       &                             & [cm$^2$\,g$^{-1}$]              & [d]            & [$10^{40}$erg\,s$^{-1}$] & [$10^3$\,K]                             & \\
  \hline
DD2-135135          & total  & 1.99e-3(1)                         & 0.255                & 0.27                                  & 1.07e-1                     & 26.5                            & 0.77           & 4.21                     & 2.75                                    & 0.36                                     \\
                    & polar  & 3.52e-4(0.18)                      & 0.252                & 0.34                                  & 4.42e-2                     & 18.3                            & 0.66           & 6.30                     & 3.24                                    & 0.30                                     \\
                    & equat. & 1.62e-3(0.82)                      & 0.256                & 0.26                                  & 1.20e-1                     & 28.3                            & 0.85           & 3.32                     & 2.50                                    & 0.38                                     \\
  \hline                                                                                                                                                                                                                                                                                                   
DD2-125145          & total  & 3.20e-3(1)                         & 0.248                & 0.22                                  & 1.52e-1                     & 29.1                            & 1.04           & 4.12                     & 2.39                                    & 0.34 \\
                    & polar  & 5.01e-4(0.16)                      & 0.255                & 0.29                                  & 6.60e-2                     & 22.3                            & 0.88           & 5.98                     & 2.98                                    & 0.28 \\
                    & equat. & 2.66e-3(0.84)                      & 0.247                & 0.21                                  & 1.68e-1                     & 30.4                            & 1.15           & 3.35                     & 2.19                                    & 0.35 \\
  \hline                                                                                                                                                                                                                                                                                                   
SFHo-135135         & total  & 3.29e-3(1)                         & 0.302                & 0.26                                  & 1.15e-1                     & 27.2                            & 0.91           & 5.34                     & 2.49                                    & 0.43\\
                    & polar  & 5.49e-4(0.17)                      & 0.241                & 0.30                                  & 8.15e-2                     & 24.6                            & 0.78           & 7.53                     & 2.96                                    & 0.33 \\
                    & equat. & 2.70e-3(0.83)                      & 0.314                & 0.25                                  & 1.22e-1                     & 27.8                            & 0.99           & 4.24                     & 2.26                                    & 0.46 \\
  \hline                                                                                                                                                                                                                                                                                                   
SFHo-125145         & total  & 8.67e-3(1)                         & 0.241                & 0.24                                  & 1.14e-1                     & 27.4                            & 1.49           & 7.51                     & 2.20                                    & 0.37 \\
                    & polar  & 1.77e-3(0.20)                      & 0.225                & 0.28                                  & 6.62e-2                     & 23.2                            & 1.33           & 9.64                     & 2.52                                    & 0.33 \\
                    & equat. & 6.75e-3(0.80)                      & 0.244                & 0.23                                  & 1.27e-1                     & 28.5                            & 1.57           & 6.72                     & 2.09                                    & 0.38 \\
  \hline                                                                                                                                                                                                                                                                                                   
DD2-135135-noneu    & total  & 1.99e-3(1)                         & 0.255                & 0.13                                  & 2.46e-1                     & 32.3                            & 0.86           & 2.79                     & 2.35                                    & 0.35 \\
                    & polar  & 3.53e-4(0.18)                      & 0.252                & 0.17                                  & 2.22e-1                     & 32.0                            & 0.73           & 3.80                     & 2.67                                    & 0.30 \\
                    & equat. & 1.62e-3(0.82)                      & 0.256                & 0.12                                  & 2.51e-1                     & 32.4                            & 0.96           & 2.30                     & 2.20                                    & 0.36 \\
  \hline                                                                                                                                                                                                                                                                                                   
DD2-135135-Ye$-$01  & total  & 1.99e-3(1)                         & 0.255                & 0.17                                  & 2.02e-1                     & 30.7                            & 0.85           & 3.18                     & 2.47                                    & 0.35 \\
                    & polar  & 3.53e-4(0.18)                      & 0.252                & 0.24                                  & 1.37e-1                     & 27.6                            & 0.70           & 4.69                     & 3.04                                    & 0.29 \\
                    & equat. & 1.62e-3(0.82)                      & 0.256                & 0.16                                  & 2.16e-1                     & 31.4                            & 9.48           & 2.55                     & 2.25                                    & 0.36 \\
  \hline                                                                                                                                                                                                                                                                                                   
DD2-135135-Ye$+$01  & total  & 1.99e-3(1)                         & 0.255                & 0.37                                  & 2.31e-2                     & 12.2                            & 0.63           & 5.09                     & 3.14                                    & 0.38  \\
                    & polar  & 3.53e-4(0.18)                      & 0.252                & 0.44                                  & 1.04e-2                     & 9.42                            & 0.53           & 7.36                     & 3.23                                    & 0.34  \\
                    & equat. & 1.62e-3(0.82)                      & 0.256                & 0.36                                  & 2.59e-2                     & 12.9                            & 0.70           & 3.88                     & 2.96                                    & 0.39 \\
\hline \hline\end{tabular}
\label{tab:prop}
\end{table*}

\subsection{Investigated models and adopted quantities}\label{sec:investigated-models}

We study the kilonova emission based on the four neutrino-hydrodynamical models of binary NS mergers that have been introduced in Part I. These models have been obtained from smoothed-particle-hydrodynamics (SPH) simulations that were performed in general relativity using the conformal flatness approximation and that employed the recently developed method ILEAS \citep{Ardevol-Pulpillo2019a} for the treatment of neutrinos. In optically thick regions, ILEAS is consistent with the diffusion law and ensures lepton-number conservation, while in optically thin regions it describes neutrino absorption using an approximation to ray-tracing. In Part I, the neutrino-hydrodynamics simulations were post-processed by extracting the hydrodynamic properties (density, entropy, and electron fraction) along the trajectories of all ejected SPH particles and computing the nucleosynthesis yields along each of these trajectories using a state-of-the-art nuclear network \citep[see, e.g.,][and references therein]{Lemaitre2021m}. In the four hydrodynamical models the nuclear EoS is varied between SFHo \citep{Steiner2013} and DD2 \citep{Typel2010}, and the two binary mass configurations 1.35\,\Msun-1.35\,\Msun and 1.25\,\Msun-1.45\,\Msun are used. We additionally consider three variations of the symmetric DD2 model, which were already introduced in Part I: In model DD2-135135-noneu we ignore all neutrino interactions that take place after the time of the merger, which we define by the time when the general relativistic lapse function reaches a first minimum. This model was denoted as the ``no neutrino'' case in Part I. Moreover, to test the impact of uncertainties in the electron fraction, $Y_e$, predicted by the neutrino-hydrodynamics simulations the two models DD2-135135-Ye$-$01 and DD2-135135-Ye$+$01 consider the cases where along all trajectories $Y_e$ was artificially decreased or increased by 0.1, respectively. Table~\ref{tab:prop} summarizes the ejecta masses, velocities, average electron fractions, and average lanthanide mass fractions for all models. See Part I for more details regarding the underlying neutrino-hydrodynamics simulations, method of ejecta extraction, employed nuclear network, and the nucleosynthesis results of all models.

\subsection{Data mapping}\label{sec:data-mapping}

Ideally, we would conduct the hydrodynamical simulations of the ejecta until they reach homology, i.e. until the thermal pressure becomes dynamically irrelevant, mixing ceases, and the velocities of Lagrangian fluid elements freeze out. For this study, however, we avoid the significant computational efforts required to follow the long-term expansion and simply assume that homology is already reached at the times $t_{\rm hyd}$ when the hydrodynamical simulations have been stopped, where $t_{\rm hyd}\approx 10-20\,$ms after the NSs fall into each other. While this assumption is likely to be justified for the high-velocity component of the dynamical ejecta, it may be less appropriate for the slower outflow particles. At $t=t_{\rm hyd}$ roughly $10-20\,\%$ of the total ejecta energy still resides in thermal energy, suggesting that the average velocities could still increase by about $5-10\,\%$. We note that our approach of fixing the velocities early after the merger also implies that we cannot describe fallback of gravitationally bound material \citep[e.g.][]{Fernandez2015c, Ishizaki2021o} or the dynamical impact of radioactive r-process heating \citep[e.g.][]{Rosswog2014, Klion2021y}.

Our method of computing the kilonova adopts as input the hydrodynamic properties and nucleosynthesis data of exactly the same outflow trajectories as discussed in Part I. Homologous expansion means that the radial coordinate, $r$, can be replaced by the radial velocity, $v$, because $v(r,t)=r/t$ holds at any given time $t$. Since our kilonova solver operates on an axisymmetric finite-volume grid, we first need to map all required data from particle trajectories, along which the nucleosynthesis calculations have been performed, onto a 2D spherical polar grid that is spanned in velocity space by the normalized radial coordinate
\begin{align}\label{eq:xdef}
  x\equiv \frac{r}{ct}
\end{align}
(with the speed of light $c$) and polar angle, $\theta$. The coordinates of each trajectory particle in this velocity space are given by its radial velocity and the polar angle\footnote{The center and orientation of the axisymmetric spherical polar coordinate system is uniquely defined by the vector of the total orbital angular momentum of the stellar binary and the center of mass. As usual, the polar angle, $\theta$, is the angle between the coordinate vector and the north pole.} along which it is ejected, both measured at $t=\thyd$. For our calculations we ignore ejecta in the southern hemisphere and assume equatorial symmetry, which is found to be rather well fulfilled as shown in Part I. Having identified the position of each ejecta particle in axisymmetric velocity space, we can now interpolate the quantities needed for the upcoming kilonova evolution to all cells of the 2D velocity grid. We map altogether five quantities, namely the mass density, $\rho$, the electron fraction at the onset of network calculations, $Y_e(\rho_{\rm nuc})$, the specific heating rate, $Q_{\rm heat}(t)$ (defined in Sect.~\ref{sec:heating-rate-opacity}), the lanthanide mass fraction\footnote{As in Part I, we subsume both lanthanides and actinides under what we call lanthanide fraction here.}, $X_{\mathrm{LA}}(t)$, as well as the average baryon mass, $\langle A_{\rm nuc}\rangle(t)$. Note that $Y_e(\rho_{\rm nuc})$ is extracted merely for diagnostic purposes (e.g. Fig.~\ref{fig_cont}) but otherwise not needed by our kilonova scheme, because the dependence of the kilonova on the electron fraction is already included in the quantities $Q_{\rm heat}$, $X_{\mathrm{LA}}$, and $\langle A_{\rm nuc}\rangle$ provided by the nucleosynthesis calculations. While $\rho$ and $Y_e(\rho_{\rm nuc})$ only need to be mapped once\footnote{For homologous expansion, the density at any given time can be obtained using $\rho\propto t^{-3}$.} the remaining quantities are time dependent and therefore need to be mapped for a sufficiently large number ($\sim 1000$ in our case) of discrete times, between which linear interpolation will be used in the upcoming evolution scheme.

For the mapping we adopt interpolation methods that are well known from SPH schemes (see, e.g., \citealp{Price2007p}). We stress, however, that our mapping procedure is completely independent of the SPH code that was used to perform the hydrodynamical simulations. Thus, the kilonova scheme presented here can equally well be applied in cases where the outflow trajectories have been extracted from grid-based simulations. For a set of $N$ outflow particles with masses $m_j$ and position vectors $\vecx_j$ (where $\vecx\equiv x\mathbf{e}^r(\theta)$ with radial unit vector $\mathbf{e}^r$ and $j=1,\ldots, N$), we obtain the density at an arbitrary position $\vecx$ using
\begin{align}\label{eq:rhointerpol}
  \rho(\mathbf{x}) = \frac{1}{2\pi \tilde{R}}\sum_j m_j W_{\mathrm{2D}}(\mathbf{x}-\mathbf{x}_j,h_j) \, ,
\end{align}
while all other quantities, represented by $A$, are interpolated as
\begin{align}\label{eq:Ainterpol}
  A(\mathbf{x}) = \frac{\sum_j^N \frac{m_j}{\rho_{\mathrm{2D},j}} A_j W_{\mathrm{2D}}(\mathbf{x}-\mathbf{x}_j,h_j)}
  {\sum_j^N \frac{m_j}{\rho_{\mathrm{2D},j}} W_{\mathrm{2D}}(\mathbf{x}-\mathbf{x}_j,h_j)} \, ,
\end{align}
from the corresponding values at the particle positions, $A_j$. In the above equations, $W_{\rm 2D}$ is the two-dimensional cubic spline kernel \citep[see, e.g.,][]{Monaghan1992y}, $h_j$ is the smoothing length,
\begin{align}\label{eq:rho2Dinterpol}
  \rho_{\mathrm{2D},j} = \sum_i^N m_i W_{\mathrm{2D}}(\mathbf{x}_j-\mathbf{x}_i,h_i)
\end{align}
is the particle representation of the 2D density, $\rho_{\mathrm{2D}} \equiv 2\pi R\rho$ \citep[e.g.][]{Garcia-Senz2009x}, $R$ is the cylindrical radius in normalized velocity space, and $\tilde{R}=\max\{R,h/2\}$ with $h$ interpolated from $h_j$ using Eq.~(\ref{eq:Ainterpol}). As commonly done in SPH schemes, we fix the smoothing length, $h_j$, by the condition that any sphere of radius $h_j$ around particle $j$ should contain roughly the same number of particles \citep[$\sim 50$ in our case; see, e.g.,][for explicit equations]{Price2007p}. In order to reduce numerical artefacts close to the polar axis, namely  at cylindrical radii $R\ll h$ that are unresolved by the particle data, we limit the conversion factor $(2\pi R)^{-1}$ between $\rho_{\mathrm{2D}}$ and $\rho$ from below by replacing $R$ with $\tilde{R}$ in Eq.~(\ref{eq:rhointerpol}).

\subsection{Evolution equations}\label{sec:lightc-calc}

After mapping the particle data onto the grid, we compute the kilonova light curves in a simplified fashion using a truncated two-moment scheme with analytic closure (or often called M1-scheme; e.g. \citealp{Minerbo1978, Levermore1984a, Audit2002a, Just2015b}). Instead of evolving the specific intensity $I_\nu(\vecx,\vecn)$ as function of the photon frequency $\nu$ and photon momentum unit vector $\vecn$, as is done in full-fledged radiative transfer schemes \citep[e.g.][]{Kasen2006w, Tanaka2013}, our approximate scheme evolves the 0th and 1st angular moments of $I_\nu$ as measured in the comoving (or fluid) frame, namely the monochromatic energy density,
\begin{align}\label{eq:0thmom}
  E_\nu(\vecx) \equiv \frac{1}{c}\int I_\nu \dd \Omega \, ,
\end{align}
and the monochromatic energy-flux density vector,
\begin{align}\label{eq:1stmom}
  F^i_\nu(\vecx) \equiv \int I_\nu\,n^i \dd \Omega \, ,
\end{align}
where $\int\dd \Omega$ denotes angular integration over the full sphere in photon momentum space. The $\mathcal{O}(v/c)$ evolution equations for these quantities as function of the velocity coordinate vector $\vecx$ can be derived from the ordinary two-moment equations \citep[e.g.][]{Mihalas1984, Just2015b} by making use of the homology condition, $v(r,t)=r/t$, and they are given by (see, e.g., \citealp{Pinto2000a, Rosswog2018a} for an analog derivation of the 0th-moment equation):
\begin{subequations}\label{eq:m1eqs}
  \begin{align}
    \frac{\dd E_\nu}{\dd t} + \frac{1}{ct}\nabla_\vecx \cdot \vecF_\nu + \frac{4E_\nu}{t}-\frac{1}{t}\frac{\partial}{\partial\nu}(\nu E_\nu) & = c\rho\kappa (E^{\mathrm{eq}}_\nu - E_\nu) \, , \label{eq:m0evo} \\
    \frac{\dd \vecF_\nu}{\dd t} + \frac{c}{t}\nabla_\vecx \cdot \tenP_\nu + \frac{4\vecF_\nu}{t}-\frac{1}{t}\frac{\partial}{\partial\nu}(\nu \vecF_\nu) & = -c\rho\kappa \vecF_\nu  \, ,
  \end{align}
\end{subequations}
where $\kappa$ is the absorption opacity (see Sect.~\ref{sec:heating-rate-opacity} for the computation), $E^{\mathrm{eq}}_\nu$ is the equilibrium (Bose-Einstein) energy density, and $\tenP_\nu/E_\nu$ is the normalized 2nd moment (Eddington) tensor with the components
\begin{align}\label{eq:eddequ}
  \frac{P^{ij}_\nu}{E_\nu} \equiv \frac{1}{c E_\nu} \int I_\nu\,n^i n^j \dd \Omega \, .
\end{align}
The M1 approximation consists of assuming that the Eddington tensor is given in terms of a local closure relation as function of $E_\nu$ and $\vecF_\nu$. We employ the closure relation by \citet{Minerbo1978}. In Eqs.~(\ref{eq:m1eqs}) the time derivative $\dd/\dd t$ is taken at constant velocity, $\mathbf{x}$, and the spatial derivatives, $\nabla_{\vecx}$, are taken with respect to $\vecx$. The individual terms entering the time derivative in Eqs.~(\ref{eq:m1eqs}) describe, from left to right, the propagation of radiation fluxes, losses due to expansion, Doppler-shift, and (emission and absorption) interactions with ions and electrons.

The energy equation of photons (cf. Eq.~(\ref{eq:m0evo})), is coupled to the energy equation for the remaining particles -- which will collectively be denoted as gas or fluid in this work -- via the first law of thermodynamics for Lagrangian fluid elements moving with velocity $v=r/t$, 
\begin{align}\label{eq:gasequ}
  \frac{\dd e}{\dd t} +\frac{5e}{t} = \rho Q_{\rm heat} - \int_0^\infty c\rho\kappa (E^{\mathrm{eq}}_\nu - E_\nu)\dd \nu \, ,
\end{align}
where $e$ is the thermal energy density of the gas. Equation~(\ref{eq:gasequ}) takes into account $p\dd V$ expansion work, energy input from radioactive decay of freshly synthesized elements that powers the kilonova (see Sect.~\ref{sec:heating-rate-opacity} for its computation), and exchange of energy due to emission and absorption of radiation. Since by the time $t>\tkn$, where $0.01\,\mathrm{d}\la \tkn\la 100\,$d are typical kilonova emission timescales, most electrons are recombined and free neutrons, protons, and positrons have disappeared, the EoS of the gas is dominated by heavy ions and given by
\begin{align}\label{eq:gaseos}
  e = \frac{3\rho k_B T}{2 \langle A_{\mathrm{nuc}}\rangle m_u} \, ,
\end{align}
where $T$, $k_B$, and $m_u$ are the fluid temperature, Boltzmann constant, and atomic mass unit, respectively, and the average mass number of ions, $\langle A_{\mathrm{nuc}}\rangle$, is provided by the nucleosynthesis calculations.

The evolution during the intermediate phase between the merger and the kilonova emission, namely during $\thyd<t<\tkn$, is less important, because in this adiabatic phase the total (photon plus gas) energy quickly converges towards a time-dependent quasi-equilibrium that is determined by the balance between adiabatic expansion and radioactive heating. Hence, as long as the time of initialization, $t_0$, is chosen to be early enough for the system to reach the quasi-equilibrium well before $t\approx\tkn$, the resulting light curve should be insensitive to the particular choice of initial conditions. Motivated by sensitivity tests using different values of $t_0$ and different initial gas temperatures (see Appendix~\ref{sec:sens-init-cond}) we initiate our simulations at $t=100\,$s after the merger using as initial conditions a homogeneous temperature of $T=100\,$K and negligibly small radiation energies. The duration of our kilonova simulations is constrained by the fact that we assume local thermodynamic equilibrium (LTE), which implies that our scheme is not applicable at late times, typically beyond $t\sim 5-20\,$d (depending on the ejecta properties) when non-LTE effects become dominant \citep[e.g.][]{Waxman2019h}. We furthermore note that the possibility of a neutron precursor \citep{Metzger2015}, which could be produced through the decay of free neutrons in the fastest ($v/c\ga 0.5$) layers of the ejecta, is not discussed in this study, mainly because such a signal cannot properly be described by our $\mathcal{O}(v/c)$ scheme.

The numerical methods employed to solve the M1 equations, of which the canonical form can be recovered by rescaling the time coordinate as shown in Appendix B of \citet{Just2021a}, are detailed in \citet{Just2015b}. We employ the same code ALCAR that is described in \citet{Just2015b} and was previously used to evolve the M1 equations for neutrino transport.

A few comments are in order regarding the advantages and disadvantages of the M1 scheme. While the M1 scheme has been employed already in a large number of previous applications in the context of photon and neutrino transport \citep[e.g.][]{Cernohorsky1994, Smit2000, Pons2000, Audit2002a, McKinney2014a, Just2015b, OConnor2015a, Kuroda2016, Weih2020a} the present scheme is, to our knowledge, the first application to the computation of kilonovae. A compelling advantage of the M1 scheme is its computational efficiency and algorithmic simplicity compared to full scale radiative transfer schemes that resolve the angular distribution of the photon field. The hyperbolic nature of the M1 equations allows to integrate spatial derivatives explicitly in time and by that avoids inversions of large matrices during each integration step \citep[e.g.][]{Just2015b}, keeping the computational expense comparable to that of hydrodynamic solvers. The accuracy of the scheme is generally expected to be superior to some widely employed, more approximate methods. For instance, the leakage-like model by \citet[][]{Grossman2014} does not solve a conservation equation for the photon energy, and as a consequence it systematically underestimates the luminosities around peak epoch\footnote{This is because the luminosity estimated in the model by \citet{Grossman2014} is always bound to be lower than or equal to the current global heating rate, which is inconsistent with more detailed calculations where the luminosity typically exceeds the heating rate at times close to the peak. We refer the reader to the Appendix of \citet{Rosswog2018a} for a comparison of the model by \citet[][]{Grossman2014} with more accurate schemes.}. In contrast to (quasi-)one-zone approximations that do solve an energy conservation equation for photons \citep[e.g.][]{Arnett1982, Goriely2011, Villar2017a, Metzger2019a, Hotokezaka2020a}, the M1 formulation does not depend on manually chosen estimates of the diffusion rate, because it self-consistently resolves the spatial propagation of radiation through the ejecta. Moreover, it is able to consistently describe heating due to reabsorption of photons transported from one location to another, and it provides a genuinely multi-dimensional framework that can handle non-radial fluxes.

On the other hand, the main disadvantage of the M1 scheme is its poor ability to describe radiation in the optically thin regime, particularly in the case of geometrically complex radiation sources \citep[e.g.][]{Audit2002a, Just2015b, Weih2020a}. While the underlying Boltzmann equation predicts linear superposition of radiation packets in the optically thin regime, the non-linear closure relation of the M1 scheme causes radiation beams to interact with each other even where collisional interaction rates vanish. A particularly noteworthy consequence of this shortcoming is that radiation emitted in the optically thin phase of evolution, after the photosphere has disappeared, is not isotropic, as it should be, but still retains a certain dependence on the observer angle, $\theta_{\rm obs}$ (cf. Sect.~\ref{sec:light-curv-fiduc}). In view of the approximate handling in M1 of the angular photon distribution function, we consider the obtained dependence of radiation fluxes on specific observer angles to be less reliable than corresponding averages over finite angle intervals. This is why we will restrict most of our discussion to considering luminosities averaged over finite solid angle domains instead of ones measured at specific angles.

\subsection{Heating rate and opacity}\label{sec:heating-rate-opacity}

\begin{figure*}
\includegraphics[width=0.99\textwidth]{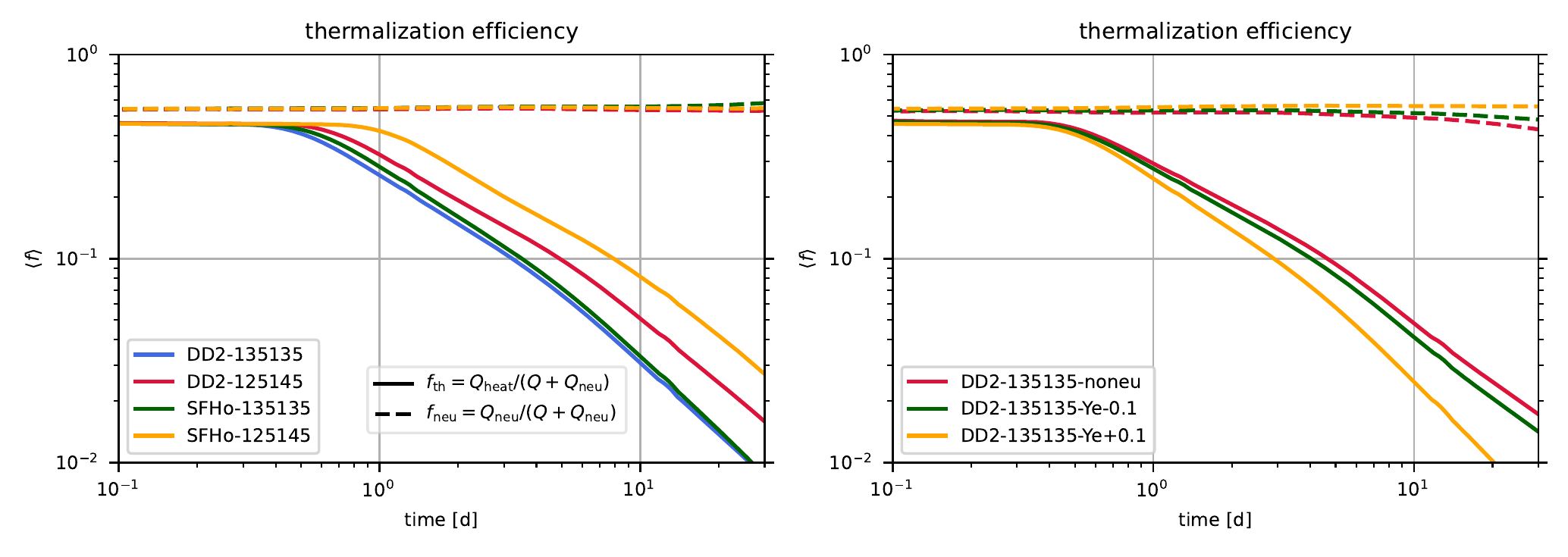}
\caption{Mass-averaged thermalization efficiencies as function of time for all of our models as computed with Eq.~(\ref{eq:fth}) \emph{(solid lines)} as well as the mass-averaged fraction of the radioactive energy release rate that is carried away by neutrinos \emph{(dashed lines)}.}
\label{fig_ftherm}
\end{figure*}

The quantity $\Qheat$ on the right-hand side of Eq.~(\ref{eq:gasequ}) is computed as
\begin{align}\label{eq:kappa}  
  \Qheat \equiv \fth (Q+\Qneu)
\end{align}
and represents the effective heating rate per unit of mass, namely the fraction $\fth$ of the total radioactive energy release rate, $Q+\Qneu$, that is converted to thermal energy on timescales shorter than the evolution timescale. We follow the notation of Part I and use $Q$ to denote the release rate without neutrinos, while $\Qneu$ stands for the contribution that is carried away by neutrinos (and not available for thermal heating in any case). We compute the thermalization efficiency, $\fth$, exactly as in \citet{Rosswog2017b} (who adopted the formalism by \citealp{Barnes2016a}), namely as
\begin{align}\label{eq:fth}  
  \fth = \frac{Q_\beta\left[\tilde\zeta_\gamma f_\gamma+\tilde\zeta_e f_e\right]+Q_\alpha f_\alpha + Q_{\rm fis} f_{\rm fis}}{Q+\Qneu} \, ,
\end{align}
where $Q_{\beta}$, $Q_{\alpha}$, and $Q_{\mathrm{fis}}$ are the partial release rates\footnote{See Sect.~3.2 and Fig.~9 in Part I for a discussion of the individual contributions to $Q$.} from $\beta$-decay, $\alpha$-decay, and fission, and the thermalization efficiencies $f_i$ ($i\in\{\gamma, e, \alpha, \mathrm{fis}\}$) for photons, electrons, $\alpha$ particles, and fission products are functions of $\tau_i/t$ with the thermalization timescales $\tau_i$; see \citet{Rosswog2017b} for the explicit expressions. Since in our notation $Q_\beta$ does not contain the energy release going into neutrinos, the weighting factors $\tilde\zeta_\gamma=0.69$ and $\tilde\zeta_e=0.31$ differ from the ones ($\zeta_\gamma=0.45$ and $\zeta_e=0.2$) employed in \citet{Rosswog2017b}. Note that while the timescales $\tau_i$ are functions only of the total mass and average velocity of the ejecta, $\fth$ can be different along each trajectory because of different relative contributions of $Q_{\beta}$, $Q_{\alpha}$, and $Q_{\mathrm{fis}}$. The average thermalization efficiencies for all investigated models as functions of time are displayed in Fig.~\ref{fig_ftherm}. Almost perfect thermalization ($\Qheat\approx Q$) prevails until about $t\sim 0.5-1\,$d, whereafter $\fth$ declines roughly as $\propto t^{-1}$. Lower ejecta masses or faster expansion velocities accelerate the decay of $\fth$. For completeness we also plot the fraction $f_{\rm neu}\equiv \Qneu/(Q+\Qneu)$ of energy that is carried away by neutrinos, which amounts to about $\approx 50-60\,\%$ in our models.

Finally, we need a reasonable prescription for the opacities, which are dominated mainly by thousands of possible transition lines between energy levels of $f$-shell elements newly created by the r-process. Instead of using a detailed atomic-physics based opacity model \citep[e.g.][]{Kasen2013, Tanaka2020a, Wollaeger2018a} in our approximate kilonova solver, we express the opacity in a parametric fashion as a function of the lanthanide mass fraction and gas temperature as
\begin{align}\label{eq:kappatot}  
  \kappa(\XLA,T) = &  \: \kappa_{\rm LA} \times \kappa_{T}
\end{align}
where the $\XLA$-dependent part is
\begin{align}\label{eq:kappala}  
  \kappa_{\rm LA} \equiv \begin{cases}
    30 \,\mathrm{cm}^{2}\,\mathrm{g}^{-1} (\XLA/10^{-1})^{0.1} & \text{, $\XLA>10^{-1}$} \, ,\\
    3 \,\mathrm{cm}^{2}\,\mathrm{g}^{-1} (\XLA/10^{-3})^{0.5} & \text{, $10^{-3}<\XLA<10^{-1}$} \, ,\\
    3 \,\mathrm{cm}^{2}\,\mathrm{g}^{-1} (\XLA/10^{-3})^{0.3} & \text{, $10^{-7}<\XLA<10^{-3}$} \, ,\\
    0.2 \,\mathrm{cm}^{2}\,\mathrm{g}^{-1} & \text{, $\XLA<10^{-7}$} \, ,
\end{cases}
\end{align}
and the temperature-dependent part is
\begin{align}\label{eq:kappatem}  
  \kappa_T \equiv \begin{cases}
    1 & \text{, $T>2000\,$K} \\
    \left(\frac{T}{2000\,\mathrm{K}}\right)^5 & \text{, $T<2000\,$K} \, .
    \end{cases}
\end{align}
The above prescription was motivated by fits to bolometric light curves from the atomic-physics based models of \citet{Kasen2017a}; see Appendix~\ref{sec:opacity-calibration} for comparison plots. Our simplified opacities are not able to reproduce all quantitative features in all models by \citet{Kasen2017a}, but they produce agreement in the bolometric luminosities typically within a factor of two, and they rather reliably predict the correct frequency regime and epoch of maximum brightness of individual broadband light curves. While our prescription is only a crude approximation to atomic-physics based models, it is more advanced than many previous treatments using (piecewise) constant opacities \citep[e.g.][]{Villar2017a, Perego2017a, Rosswog2018a, Hotokezaka2020a} in two important respects: First, the functional dependence between $\XLA$ and $\kappa$ is more consistent with sophisticated opacities. In particular, it ensures that variations in the lanthanide fractions always result in variations of the opacity and therefore of the light curve, which is not guaranteed for a piecewise constant prescription. Second, the temperature-dependent reduction factor, $\kappa_T$, incorporates the effect that electron recombination and the concomitant disappearance of atomic transition lines lead to a decline of the opacities at late times \citep[e.g.][]{Kasen2013, Tanaka2020a,Zhu2021a}.

\subsection{Grid setup}\label{sec:grid-setup-init}

The hydrodynamical NS-NS merger simulations sample the outflow in each model with $N\sim 800-1400$ trajectories (see Part I). The kilonova evolution is simulated on a velocity grid containing about 400 zones in radial direction, distributed uniformly within $v/c=0.02-1.2$\footnote{The reason for locating the outer boundary at $v>c$ is only to prevent numerical artifacts, which may be encountered at early times when radiation crosses the boundary with low values of the flux factor, $|\vecF_\nu|/(cE_\nu)$.}, and 60 equidistant zones in $\theta$ direction from $0$ to $\pi/2$. Frequency space is discretized by 50 frequency bins, which are distributed logarithmically between $\nu=0$ and $2.42\times 10^{15}\,$Hz. The midpoints of the outermost bins are located at wavelengths of $130\,$nm and $289\,\mu$m.

\subsection{Extraction of observer light curves}\label{sec:extr-observ-light}

In contrast to a complete radiative transfer solver, the approximate two-moment system with analytic closure that we solve, Eq.~(\ref{eq:m1eqs}), does not evolve the angular distribution of photons in momentum space, but only the 0th- and 1st-angular moments thereof (cf. Eqs.~(\ref{eq:0thmom}) and~(\ref{eq:1stmom})). Since the moments are defined in the frame locally comoving with the ejecta the fluxes $F^i_\nu$ measured at some velocity coordinate $v_m$ above the photosphere are Doppler red-shifted and retarded in time compared to the fluxes measured by a Eulerian observer, who is assumed to be at rest with respect to the center of mass of the NS binary. Moreover, due to the proximity of $v_m$ to spatially extended cloud of ejecta a significant fraction of photons is still moving in lateral directions, i.e. the angular distribution of $F^r_\nu$ is not the final one as measured by an observer. We apply the following scheme based on discretizing the radiation field into radiation packets in order to reconstruct the approximate angular distribution of the radiation field and to correct for the frame-dependence (inspired by a related treatment in \citealp{Lucy2005u}).

The first step consists of recording the comoving-frame radiation moments, $E_\nu$, $F^r_\nu$, and $F^\theta_\nu$ at the time-dependent velocity coordinate
\begin{align}\label{eq:vm}  
  v_m(t) = \max\{0.5c, \min\{0.99c, 1.2 \langle v_{\rm ph}\rangle\}\} \, ,
\end{align}
which is chosen to be $20\,\%$ larger than the average location of the photosphere (see Sect.~\ref{sec:light-curv-fiduc} and Eq.~(\ref{eq:vph}) for the definition of the photosphere), but within $c$ and $c/2$. The lower limit of $0.5c$ is applied in order to ensure that monitoring is performed outside of regions of significant photon production also after the ejecta become optically thin.

In the next step we reconstruct the specific intensity, $I_\nu$, using explicitly the closure prescription of \citet{Minerbo1978}. This closure assumes $I_\nu$ to be of the form
\begin{align}\label{eq:intmin}  
 I_\nu(\tilde\mu) \propto e^{a_\nu\tilde\mu}
\end{align}
as function of the cosine
\begin{align}\label{eq:mutilde}  
 \tilde\mu \equiv \vecn\cdot\frac{\vecF_\nu}{|\vecF_\nu|}
\end{align}
of the angle between the momentum vector of photons, $\vecn$, and the flux-density unit vector, $\vecF_\nu/|\vecF_\nu|$, and where the parameter $a_\nu=L^{-1}(|\vecF_\nu|/(c E_\nu))$ with the inverse of the Langevin function, $L^{-1}$. The normalization constant of $I_\nu$ is obtained from the definition of either of the evolved moments (cf. Eqs.~(\ref{eq:0thmom}) or (\ref{eq:1stmom})). Next, we discretize the momentum space of photons, using as coordinates $\mu\equiv\vecn\cdot \mathbf{e}^r \in [-1,1]$, the cosine of the angle between the photon momentum and the radial direction, and the azimuthal angle $\Phi\in [0,2\pi)$ between the projection of $\vecn$ onto the plane normal to $\mathbf{e}^r$ and the unit vector in $\theta$ direction, $\mathbf{e}^\theta$. The differential energy of a radiation packet emitted at time $t$, velocity $v=v_m(t)$, polar angle $\theta$, and frequency $\nu$ into the direction of $\vecn(\mu,\Phi)$ is then given by
\begin{align}\label{eq:epack}  
 E_{\rm pack} = I_\nu(v_m,\theta,\mu,\phi,t)\mu \dd \nu \dd\mu \dd\Phi  \times 2\pi(v_m t)^2\sin\theta\dd \theta\dd t 
\end{align}
and the number of photons carried by that packet is $N_{\rm pack}=E_{\rm pack }/(h\nu)$ (with Planck constant $h$). While $N_{\rm pack}$ is frame independent, the energy, momentum vector, and arrival time of each packet as measured by an observer at rest transform as
\begin{subequations}\label{eq:trafo}
\begin{align}
 E_{\rm pack, obs} & = E_{\rm pack} \Gamma\left(1+ \vecn\cdot \frac{\vecv}{c}\right) \, , \\
 \vecn_{\rm obs} & = \frac{E_{\rm pack}}{E_{\rm pack, obs}}\left[\vecn + \Gamma\frac{\vecv}{c}\left(1+\frac{\Gamma}{\Gamma+1}\vecn\cdot\frac{\vecv}{c}\right)\right] \, , \\
 t_{\rm obs}  & = t\left(1 - \vecn_{\rm obs} \cdot \frac{\vecv}{c}\right) \, ,
\end{align}
\end{subequations}
where $\vecv\equiv v_m \mathbf{e}^r$ and $\Gamma\equiv(1-v_m^2/c^2)^{-1/2}$. Since the time elapsed in the $v=0$ frame is the same as the observer time, we will from now on drop the subscript ``obs'' from the observer time coordinate and, for the sake of a compact presentation, use $t$ to denote both ejecta properties and the observer signal.

The differential luminosity, $\dd L_\nu$, at time $t_{\rm obs}$, observer angle $\theta_{\rm obs}$, and frequency $\nu$ is then given by the sum of energy packets $E_{\rm pack, obs}$ arriving between $t_{\rm obs}$ and $t_{\rm obs}+\dd t$ into the solid-angle interval $\dd \Omega=2\pi\sin\theta_{\rm obs}\dd \theta_{\rm obs}$ and with mean energy $E_{\rm pack, obs}/N_{\rm pack}$ between $h\nu$ and $h\nu+h\dd \nu$, divided by $\dd t$. The total luminosity going into a given solid angle is then just the sum of $\dd L_\nu$ over the appropriate angle interval. We compute the isotropic-equivalent luminosity into a given observer angle as
\begin{align}\label{eq:liso}  
 L_{\rm iso,\nu}(\theta_{\rm obs}) = 4\pi \frac{\dd L_\nu}{\dd \Omega}
\end{align}
and the flux density measured by an observer at distance $d$ as
\begin{align}\label{eq:fobs}  
 F_{\nu,\rm obs}(\theta_{\rm obs}) =  \frac{L_{\rm iso,\nu}(\theta_{\rm obs})}{4\pi d^2} \, .
\end{align}
Corresponding bolometric quantities, for which we will use the same symbol but without the subscript $\nu$, are obtained by integration over frequency. Absolute AB magnitudes are computed from the above fluxes as\footnote{By convention, absolute magnitudes assume the flux $F_{\nu,\rm obs}$ in Eq.~(\ref{eq:LAB}) to be measured at a distance of 10\,pc.}
\begin{align}\label{eq:LAB}
  M_{\rm AB} \equiv -2.5\times \log_{10}\{F_{\nu,\rm obs} [\rm{erg}\,\rm{s}^{-1}\,\rm{cm}^{-2}\,\rm{Hz}^{-1}]\} - 48.6 \, ,
\end{align}
while in this study we restrict ourselves to the three bands \textit{g, z}, and \textit{H}. The $g$ band represents green optical frequencies, the $z$ band the blue end of the near-IR domain, and the $H$ band the red end of the near-IR domain. For computing the magnitudes in these bands, we adopt the fluxes at the midpoint wavelength of each band, namely at 514, 902, and 1630\,nm, respectively, i.e. we do not apply a filter function.

Furthermore, we extract, for each observer angle, $\theobs$, and time, $t$, a temperature, $T_{\rm obs}$, and velocity, $v_{\rm obs}$, by fitting the observed isotropic-equivalent luminosities, $L_{\rm iso,\nu}$, to a blackbody luminosity, $L_{\rm iso,\nu}^{\rm BB}$, that would result if all photons were emitted with a blackbody spectrum of temperature $T_{\rm obs}$ from a spherical surface of radius $v_{\rm obs}t$, where
\begin{align}\label{eq:Lisobb}
  L_{\rm iso,\nu}^{\rm BB}(T_{\rm obs},v_{\rm obs}) \equiv\frac{8 \pi^2 h(v_{\rm obs}t)^2}{c^2}\frac{\nu^3}{\exp\{h\nu/(k_{\rm B}T_{\rm obs})\}-1}
\end{align}
with Boltzmann constant $k_{\rm B}$. The quantities $T_{\rm obs}$ and $v_{\rm obs}$ estimate the temperature and velocity of the photosphere based on observable fluxes only. In Sects.~\ref{sec:ejecta-structure} and~\ref{sec:light-curv-fiduc} we will additionally introduce the actual temperature and velocity of the photosphere, $T_{\rm ph}$ and $v_{\rm ph}$. We will provide both types of quantities for each model in order to illustrate the connection between properties of the expanding ejecta and the observable signal.

\section{Results}\label{sec:results}

In the following we consider the spatial structure of the ejecta and the photosphere in Sect.~\ref{sec:ejecta-structure}, then we examine the light curves as functions of time and observer angle for a fiducial model, and subsequently we investigate the model dependence of the results.

We summarize global properties of the ejecta and the corresponding kilonova in Table~\ref{tab:prop}. We stress that throughout this paper our definition of the kilonova peak is not strictly mathematical, i.e. we do \emph{not} identify the peak as the point where the brightness reaches a global maximum. Maximum brightness can appear already very early, in which case it is produced only by a small amount of mass located at the outer edge of the ejecta \citep[see, e.g.,][for a light-curve comparison between different choices of mass distributions in the outermost shells]{Banerjee2020b}. Since we are not interested in the emission coming from the fastest ejecta shells, but rather in that from the bulk ejecta, we measure the peak as the time when the bolometric luminosity, $L$, first starts to exceed the volume integral of the instantaneous heating rate,
\begin{align}\label{eq:qheat}
  q \equiv \int_V \rho \Qheat \dd V \, ,
\end{align}
which signals the onset of the optically thin phase\footnote{The crossing of $q$ and $L$ is a generic feature of radioactively powered transients \citep[e.g.][]{Arnett1982, Kasen2019a, Hotokezaka2020a} and a result of the fact that radioactive heating effectively charges the thermal energy content of the ejecta in the late optically thick phase. The subsequent rapid release  (called ``diffusion wave''; see, e.g., \citealp{Waxman2018a}) of this surplus of energy commencing once optically thin conditions are reached creates luminosities in excess of the instantaneous heating rates.}.

In Table~\ref{tab:prop}, as well as in the following discussions, we often distinguish between polar and equatorial properties. If not explicitly stated otherwise, ``equatorial'' always refers to the region within angles of $-\pi/4$ and $+\pi/4$ around the equator, while ``polar'' refers to the remaining volume of the sphere\footnote{We caution the reader that the two characteristic regions used here are different from the three regions (polar, middle, equatorial) used in Part I.}.

\subsection{Ejecta structure}\label{sec:ejecta-structure}

\begin{figure*}
\includegraphics[trim=0 12 0  0,clip,width=0.99\textwidth]{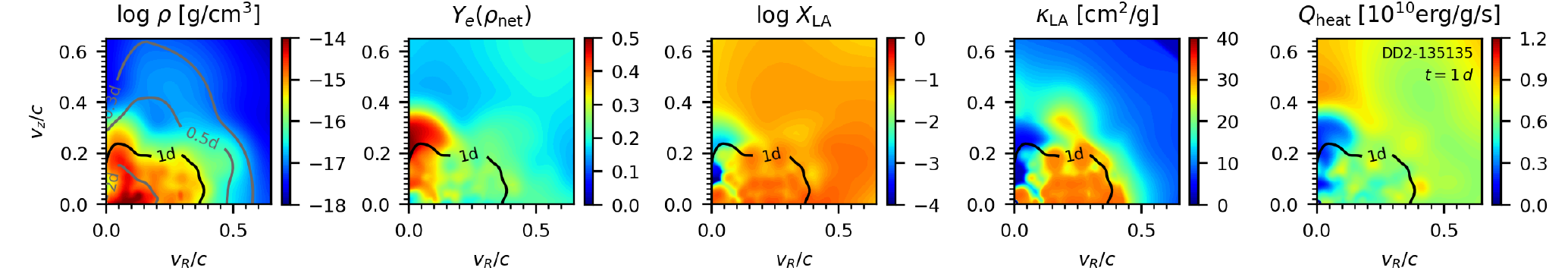}
\includegraphics[trim=0 12 0 11,clip,width=0.99\textwidth]{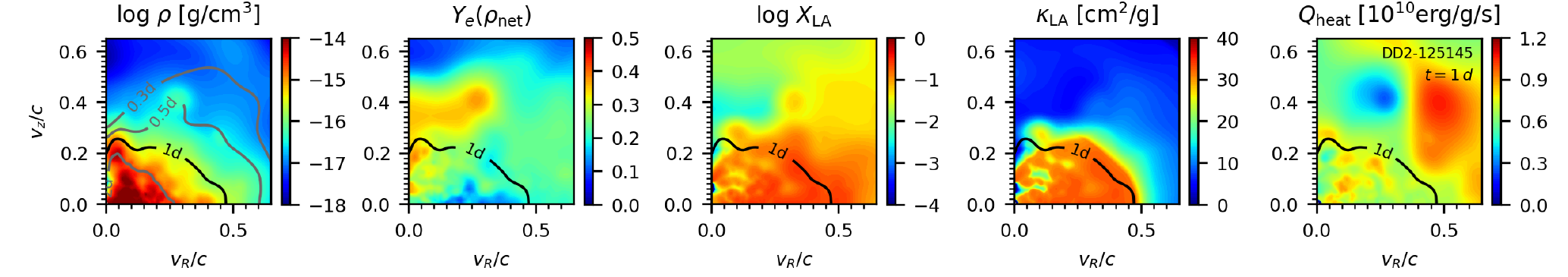}
\includegraphics[trim=0 12 0 11,clip,width=0.99\textwidth]{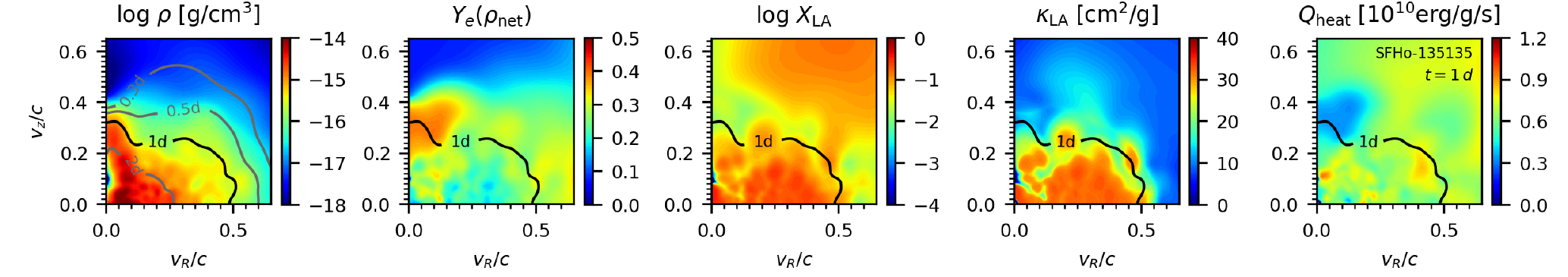}
\includegraphics[trim=0 12 0 11,clip,width=0.99\textwidth]{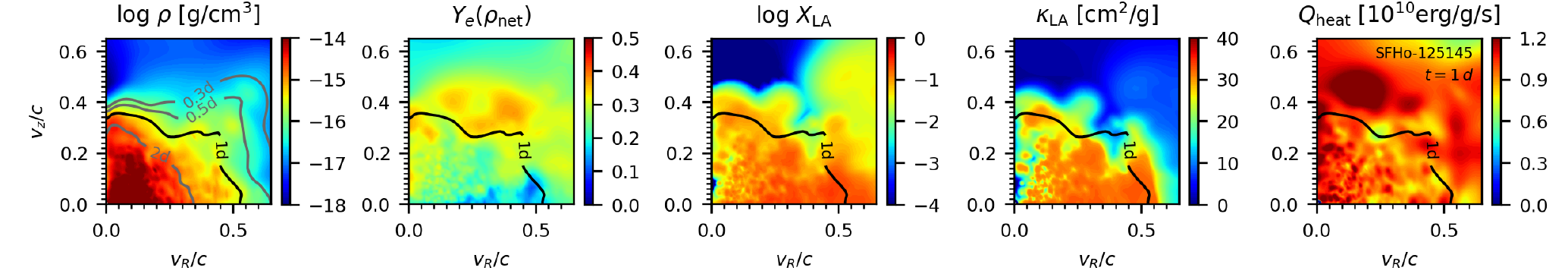}
\includegraphics[trim=0 12 0 11,clip,width=0.99\textwidth]{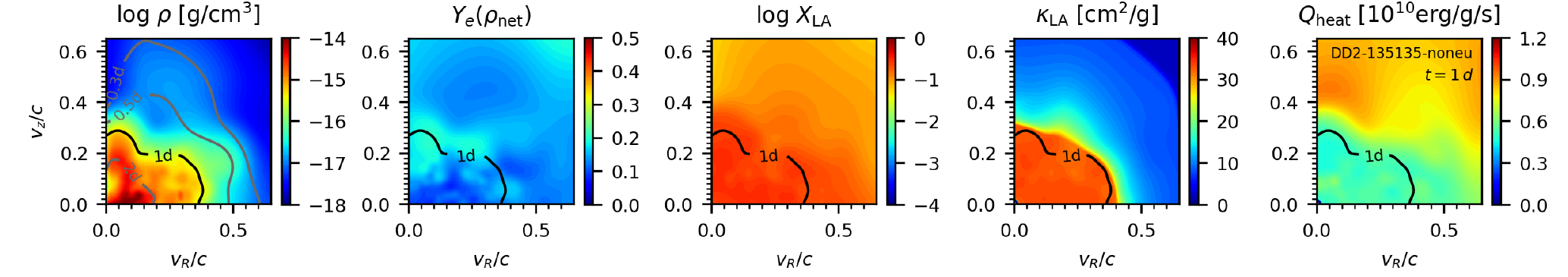}
\includegraphics[trim=0 12 0 11,clip,width=0.99\textwidth]{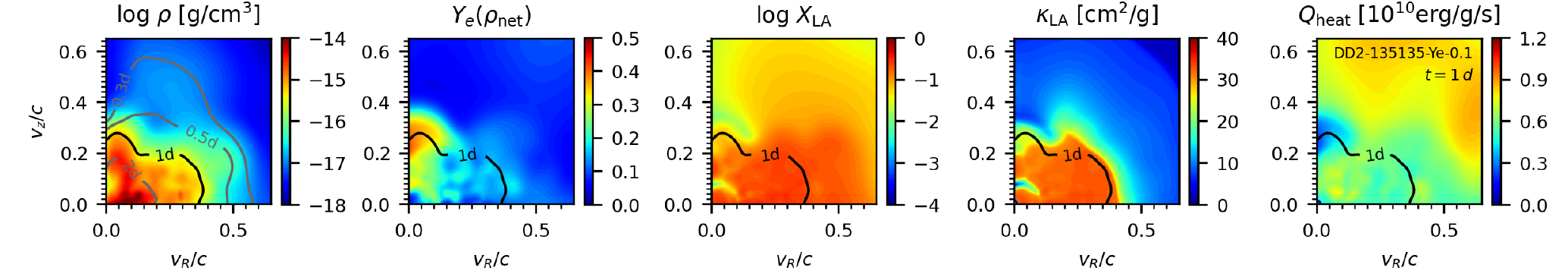}
\includegraphics[trim=0  0 0 11,clip,width=0.99\textwidth]{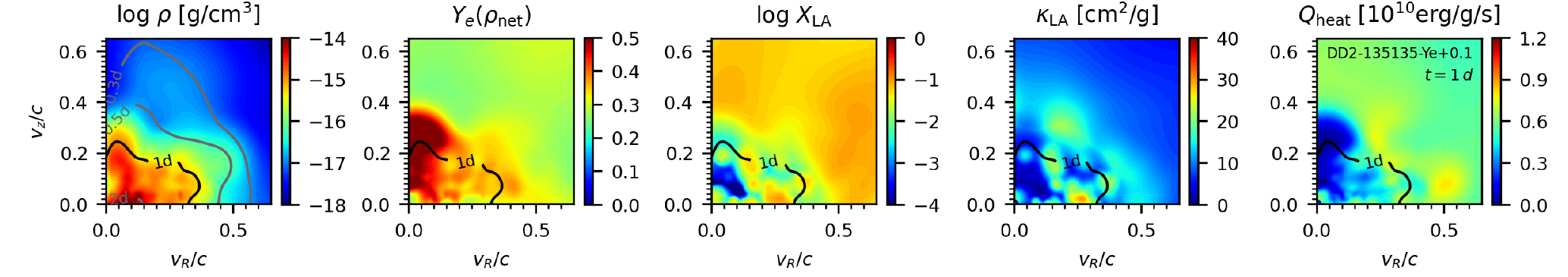}
\caption{Color maps of various quantities in velocity space at time $t=1\,$d post merger. All plots in a row belong to the model given in the right panel. From left to right the logarithmic density, electron fraction at the time when the network calculation was initiated, logarithmic lanthanide mass fraction, opacity, and specific heating rate after thermalization are shown. The black line in each panel denotes the photosphere (cf. Eq.~(\ref{eq:tau})) at $t=1\,$d, while the grey lines in the left panels additionally show the photospheres at $t=0.3, 0.5, $ and~2\,d going from high to low velocities.}
\label{fig_cont}
\end{figure*}

\begin{figure}
\includegraphics[width=0.49\textwidth]{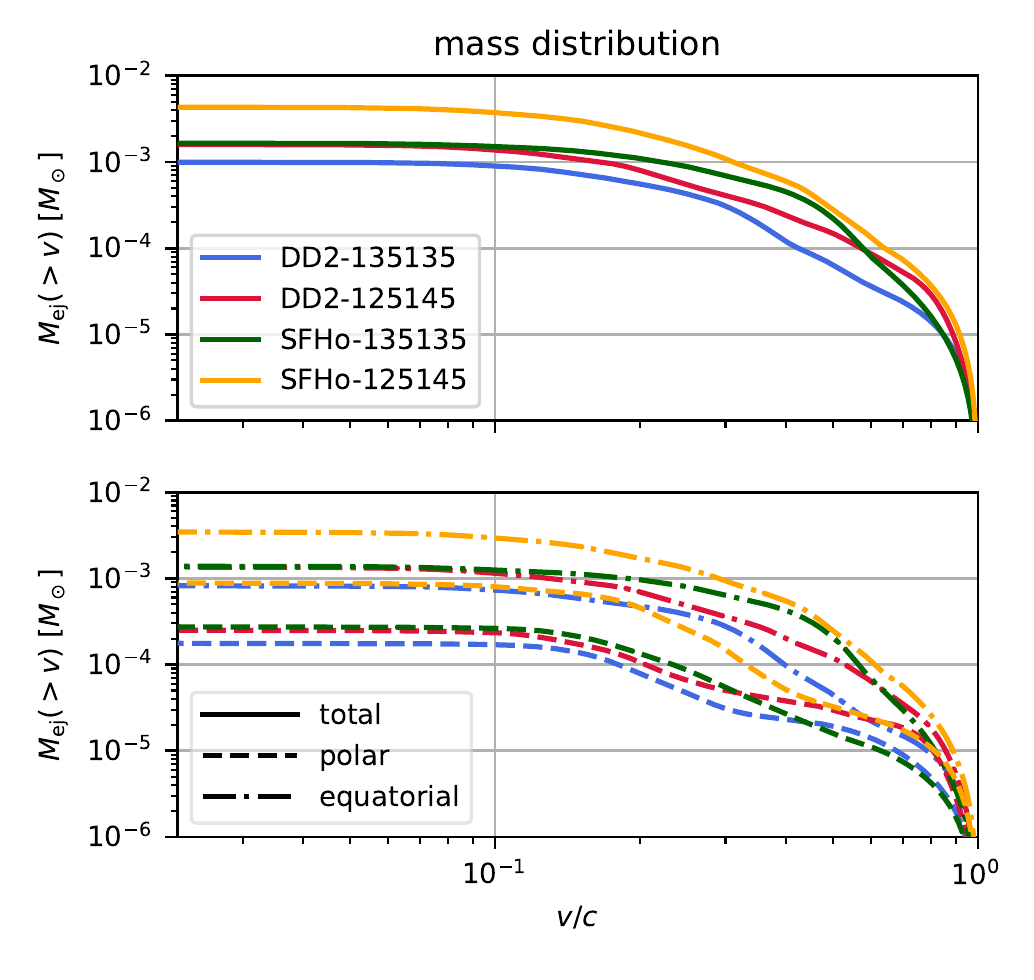}
\caption{Mass distribution of all investigated models showing for a given velocity $v$ the mass of all ejected material expanding faster than $v$. In the top panel the angle integration is performed over the entire sphere while in the bottom panel $M_{\rm ej}(>v)$ is shown separately for polar ($0<\theta<\pi/4$ and $3\pi/4<\theta<\pi$) and equatorial ($\pi/4<\theta<3\pi/4$) regions.}
\label{fig_massdis}
\end{figure}

Before considering the light curves, we first inspect the spatial distribution of the ejecta. To that end, we provide in Fig.~\ref{fig_cont} a tableau of contour plots showing for all models the density, electron fraction, lanthanide fraction, opacity, and heating rate in the polar plane at a time of $t=1\,$d post merger. Moreover, Fig.~\ref{fig_massdis} provides for all models the angle-integrated mass distributions as function of the velocity coordinate.

As already discussed in Part I, and in qualitative agreement with results by other groups \citep[][]{Sekiguchi2015a, Foucart2016a, Radice2018b}, all models show, at least for velocities $v/c\la 0.5$, a characteristic pole-to-equator asymmetry in the electron fraction, $Y_e$, which increases from $Y_e\approx 0.15-0.2$ near the equator to about $0.3-0.35$ near the poles (cf. panels in second row from left in Fig.~\ref{fig_cont}). High values of $Y_e$ beyond $\sim 0.25$ are mainly a result of neutrino absorption, which tends to dominate neutrino emission near the poles and drives $Y_e$ towards $\sim 0.5$ (e.g. \citealp{Goriely2015a}; see also Part I). As a result of the reduced neutron-richness, the mass-averaged lanthanide fraction, $\langle\XLA\rangle$, in the polar cones is about a factor of two smaller than in the equatorial region (cf. Table~\ref{tab:prop}). As a consequence, the mass-averaged opacities, $\langle\kappa_{\rm LA}\rangle$, in the corresponding regions differ by about $15-30\,\%$. Besides the composition asymmetry, polar ejecta also exhibit a slightly reduced mass loading compared to equatorial ejecta (cf. bottom panel of Fig.~\ref{fig_massdis}). The mass ratio $M_{\rm ej}^{\rm polar}/M_{\rm ej}^{\rm total}$ of polar to total ejecta is in all models about a factor of $\sim 2$ smaller than the value of 0.292 that would result for an isotropic ejecta distribution (cf. Table~\ref{tab:prop} and see Part I).

The black and grey lines in Fig.~\ref{fig_cont} denote the photosphere, which we define as the surface $r_{\rm ph}(\theta)$ where the radial optical depth, $\tau$, equals unity, i.e.
\begin{align}\label{eq:tau}
  \tau(r_{\rm ph}(\theta), \theta) = \int_{r_{\rm ph}}^\infty \kappa\rho\dd r = 1 \, .
\end{align}
Around times of $t\sim 1\,$d, i.e. close to the epoch of peak emission, the photospheres of all models resemble, to first order, an oblate ellipsoid with semi-major axis ratio of about $1.2-2$. However, the shape of the photosphere varies with time and becomes more spherical or even prolate for some models at very early or late times. Moreover, apart from the global geometry, the ejecta also exhibit a significant degree of clumpiness and small-scale structure, i.e. structures on scales of $\Delta v\sim 0.1\,c$ and less. Such variations can be a result of the case dependent merger dynamics, e.g. of mass shedding during the plunge, or oscillations of the merger remnant \citep[e.g.][]{Bauswein2013}, or of other processes such as spiral-waves \citep[e.g.][]{Nedora2019y}. As a consequence, the ejecta of all investigated models develop a photosphere with wiggles and irregular shapes and a complex time dependence (see left panel of Fig.~\ref{fig_cont} for photospheres at different times), clearly different from the idealized geometries that were assumed in previous studies \citep[e.g.][]{Kawaguchi2018a, Darbha2020b, Korobkin2021e}. The degree of clumpiness seen in our models may be exaggerated by our particle-based interpolation method and the fact that the hydrodynamics are not followed all the way until self-similar expansion, in particular because r-process heating on longer timescales may lead to smoother density profiles \citep[][]{Rosswog2014}. On the other hand, our models may also underestimate the level of anisotropy to some extent, because they are azimuthally averaged, do not include magnetic fields \citep[e.g.][]{Ciolfi2020s}, or possibly under-resolve the small-scale structure due to limited numerical resolution in the ejecta. It is well conceivable that irregular structures in the ejecta have a non-negligible bearing on the kilonova emission, e.g. by enhancing the emission anisotropy. However, since our approximate kilonova treatment is not well suited for reliably tracking the radiative transfer effects of local (in contrast to global) asymmetries in the photospheric region, we will not discuss the ramifications of the ejecta clumpiness for the kilonova lightcurve in this study, but leave such investigations to future studies based on more accurate radiative transfer solvers.

\subsection{Light curves of fiducial model}\label{sec:light-curv-fiduc}

\begin{figure*}
\includegraphics[width=0.99\textwidth]{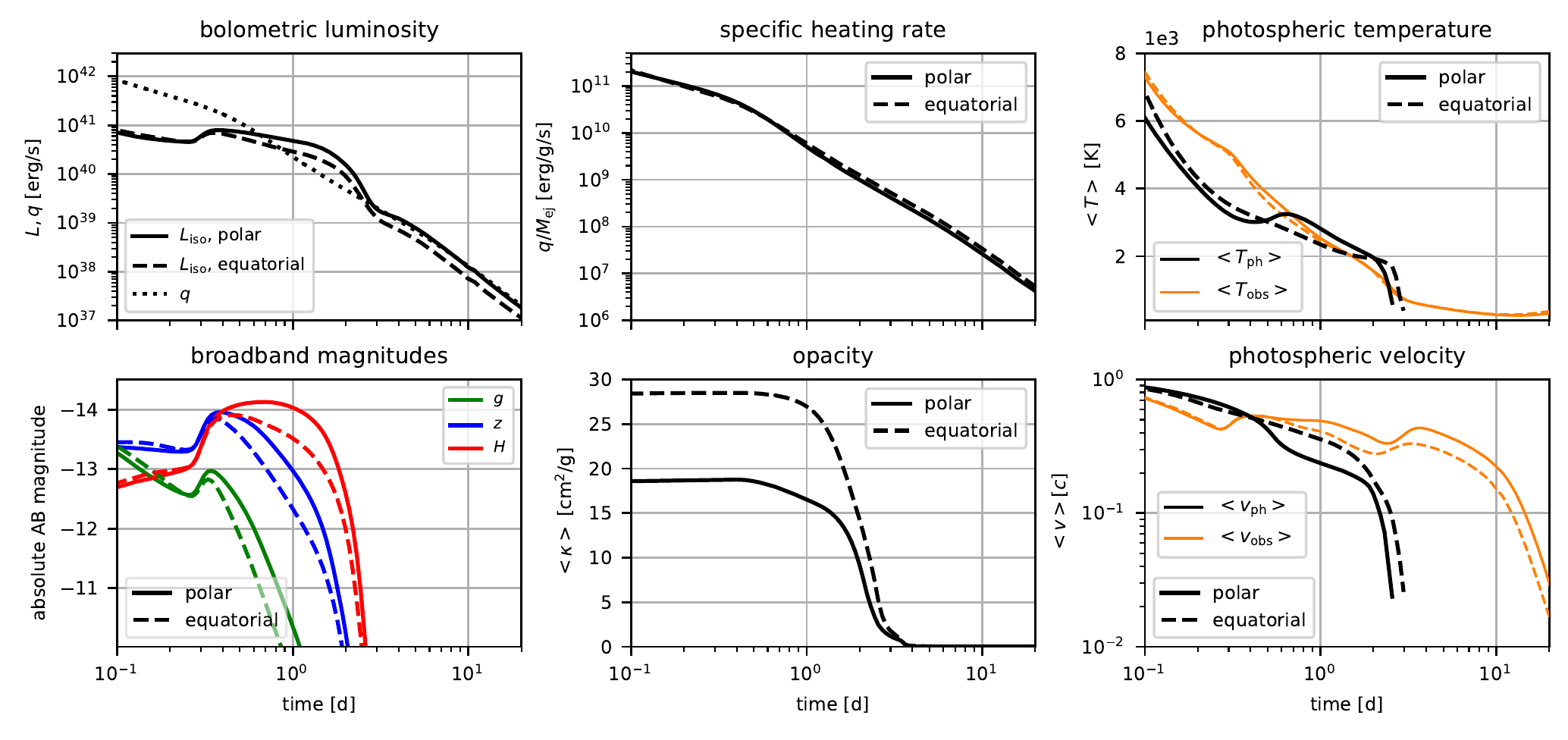}
\caption{Global properties as functions of time characterizing the kilonova for the fiducial model DD2-135135.
The left column shows in the top panel the bolometric isotropic-equivalent luminosity (cf. Eq.~(\ref{eq:liso}) together with the thermalized heating rate (cf. Eq.~(\ref{eq:qheat})) and in the bottom panel absolute AB magnitudes (cf. Eq.~(\ref{eq:LAB})), the middle column displays the mass-averaged specific heating rate (top) and mass-averaged opacity (cf. Eq.~(\ref{eq:kapp}); bottom), and the right column provides in the top panel surface averages of the photospheric temperature (cf. Eq.~(\ref{eq:Tph}); top) and photospheric velocity (cf. Eq.~(\ref{eq:vph}); bottom). Orange lines refer to the estimates of the photospheric temperature and velocity, $T_{\rm obs}$ and $v_{\rm obs}$, respectively, based on spectral fits to a blackbody spectrum (cf. Eq.~(\ref{eq:Lisobb}). Solid (dashed) lines denote volume averages ($q$, $\langle\kappa\rangle$) or surface averages (remaining quantities) over the polar (equatorial) domain. }
\label{fig_tplotpolequ}
\end{figure*}

\begin{figure}
\includegraphics[trim=0 5 0 0,clip,width=0.48\textwidth]{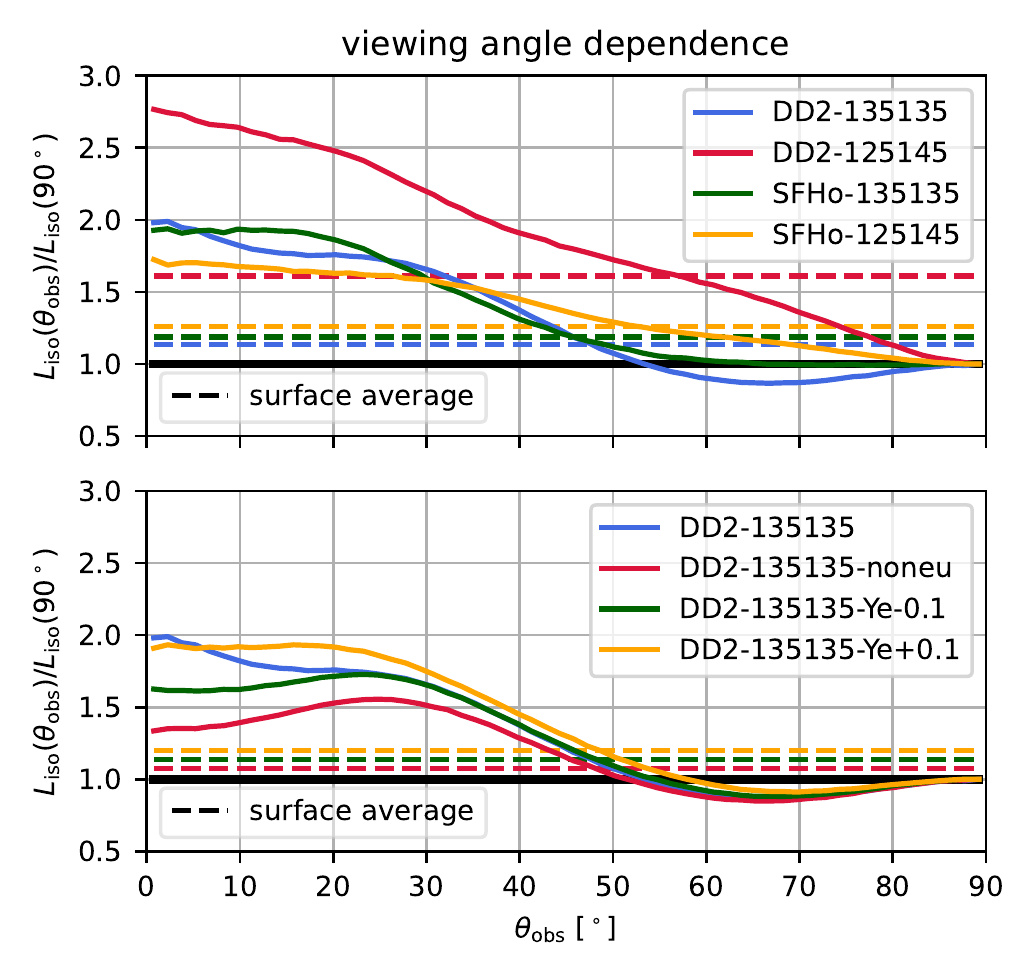}
\caption{Isotropic-equivalent luminosity as function of the inclination angle between rotation axis and observer line-of-sight, $\theta_{\rm obs}$, at $t=1\,$d normalized to the value at $\theta_{\rm obs}=90^\circ$ (solid lines). Dashed lines show the corresponding surface averages.}
\label{fig_angdep}
\end{figure}

\begin{figure}
\includegraphics[trim=0 5 0 0,clip,width=0.49\textwidth]{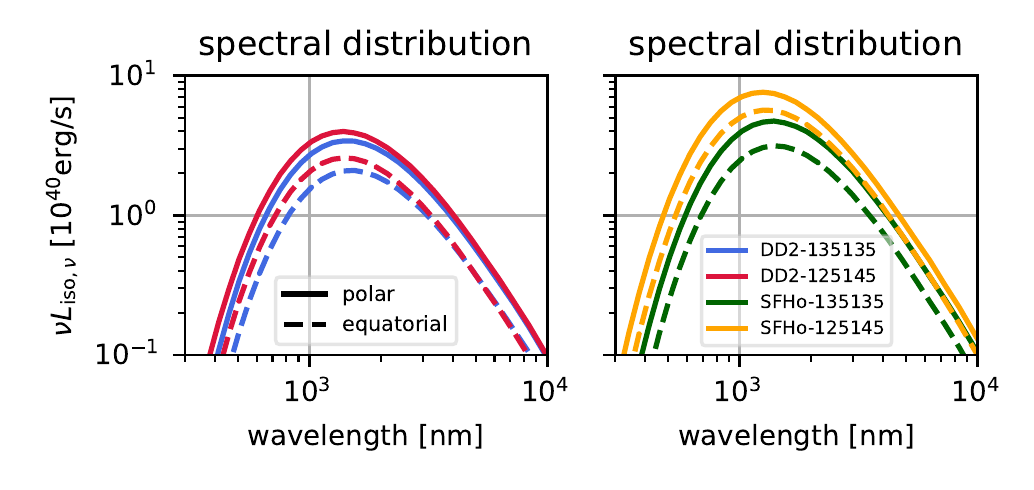}
\caption{Monochromatic, isotropic-equivalent luminosity times frequency, $\nu L_{\rm iso,\nu}$, as function of wavelength for the four models mentioned in the right panel. Solid (dashed) lines denote surface averages over the polar (equatorial) domain.}
\label{fig_spectra}
\end{figure}

We now discuss the light curve starting with the fiducial model DD2-135135. In Fig.~\ref{fig_tplotpolequ} we plot, as functions of time, the angle-averaged bolometric luminosity, the total and mass-averaged specific heating rates, $q$ and $q/\Mej$, respectively, the surface-averaged absolute magnitudes in $g$, $z$, and $H$ bands, the mass-averaged opacity,
\begin{align}\label{eq:kapp}
  \langle \kappa\rangle =  \frac{\int_V \rho\kappa \dd V}{\Mej}\, ,
\end{align}
as well as the surface-averaged temperature and velocity at the photosphere,
\begin{align}\label{eq:Tph}
  \langle T_{\rm ph}\rangle =  \frac{1}{2}\int_0^\pi T(r_{\rm ph},\theta) \sin\theta \dd\theta \, ,
\end{align}
and
\begin{align}\label{eq:vph}
  \langle v_{\rm ph}\rangle =  \frac{1}{2}\int_0^\pi \frac{r_{\rm ph}}{t} \sin\theta \dd\theta \, ,
\end{align}
respectively, together with the corresponding surface-averaged estimates $\langle T_{\rm obs}\rangle$ and $\langle v_{\rm obs}\rangle$ of the photospheric conditions based on observable fluxes.

The bulk of the ejecta starts to become optically thin at about $t\approx 0.8\,$d, which is the time when the luminosity exceeds the heating rate. By the time of $t\approx 3\,$d the optical depth has dropped below unity, i.e. the photosphere has disappeared, implying that re-absorption of thermal photons has become irrelevant.

In Fig.~\ref{fig_tplotpolequ}, the solid (dashed) lines denote quantities for which only the polar (equatorial) solid angles have been taken into account for the corresponding volume- or surface-integration. At early times of $t\la 0.2-0.3$\,d, when the photosphere is still located at high velocities of $v\ga 0.5\,c$, the composition is nearly independent of the polar angle and material is lanthanide-rich in both polar and equatorial directions (see Fig.~\ref{fig_cont}). Once the photosphere travels to slower ejecta mass shells, with $v\la 0.5\,c$, the composition gradient along the polar angle, mentioned in Sect.~\ref{sec:ejecta-structure}, starts to become relevant. The lower opacities (cf. bottom middle panel in Fig.~\ref{fig_tplotpolequ}) together with the reduced mass loading in the polar region allow the photosphere to reach lower velocities and higher temperatures in the polar compared to the equatorial ejecta for a given time. The fact that due to the angular composition gradient radiation is released from polar ejecta more readily than from equatorial ejecta, provides one reason for the excess of polar compared to equatorial values of $L_{\rm iso}$ at times $t\ga 0.3\,$d. Another reason, which is also observed for all considered models, is the oblate geometry of the photosphere at times $t\sim 1\,$d, which results in a larger projected surface area when viewed face-on compared to edge-on \citep{Darbha2020b, Wollaeger2018a}. On the other hand, angular variations of the specific radioactive heating rate do not seem to play a significant role for producing pole-to-equator emission asymmetries, as is suggested by the top middle panel in Fig.~\ref{fig_tplotpolequ} (see Part I for an in-depth discussion of the heating rate as function of polar angle and velocity).

The orange lines in the right panels of Fig.~\ref{fig_tplotpolequ} also show, for comparison to the corresponding quantities measured at the photosphere, the spectral temperature, $\langle T_{\rm obs}\rangle$, and velocity, $\langle v_{\rm obs}\rangle$, inferred from the fluxes assuming blackbody emission. The agreement between both types of quantities is good, though not perfect, at least around peak emission times of $0.7\,\mathrm{d}\la t\la 3\,\mathrm{d}$. At earlier times the spectral temperature (velocity) based on the blackbody model systematically overestimates (underestimates) the corresponding photospheric quantity. At late times, $t\ga 3\,$d, $\langle T_{\rm obs}\rangle$ and $\langle v_{\rm obs}\rangle$ do not characterize the photospheric conditions anymore, since by then the photosphere has disappeared.

Looking at magnitudes in given frequency bands (cf. bottom left panel of Fig.~\ref{fig_tplotpolequ}), the kilonova of the dynamical ejecta peaks near the red end of the near-IR band, represented by $H$ in our case, and it does so in most of our models. The $H$-band magnitude near peak epoch is, also roughly representative of all our models, $\sim -14\,$mag and about $0.5\,$mag higher for polar observers than for equatorial ones.

The dependence of $L_{\rm iso}$ on the observer angle $\theta_{\rm obs}$ is shown at $t=1\,$d in Fig.~\ref{fig_angdep}. The pole-to-equator ratio of $L_{\rm iso}$ is close to 2 for this model and similar to other models with a comparable semi-major axis ratio of the photophere (cp. black lines in Fig.~\ref{fig_cont}). Model DD2-125145 exhibits a higher value close to 3, which is likely the result of a significantly more oblate ejecta geometry compared to the other models. Finally, the behavior of the spectral distribution of $\nu L_{\rm iso,\nu}$, plotted in Fig.~\ref{fig_spectra}, does not seem to exhibit a strong dependence on the observer angle.

\subsection{Model dependence}\label{sec:model-dependence}

In the following we examine how the previously identified properties change when switching the nuclear EoS, decreasing the mass ratio of the NS binary, ignoring neutrino interactions, and artificially reducing or enhancing the electron fraction $Y_e$ by $0.1$.

\subsubsection{Equation of state and binary mass ratio}

\begin{figure*}
\includegraphics[width=0.99\textwidth]{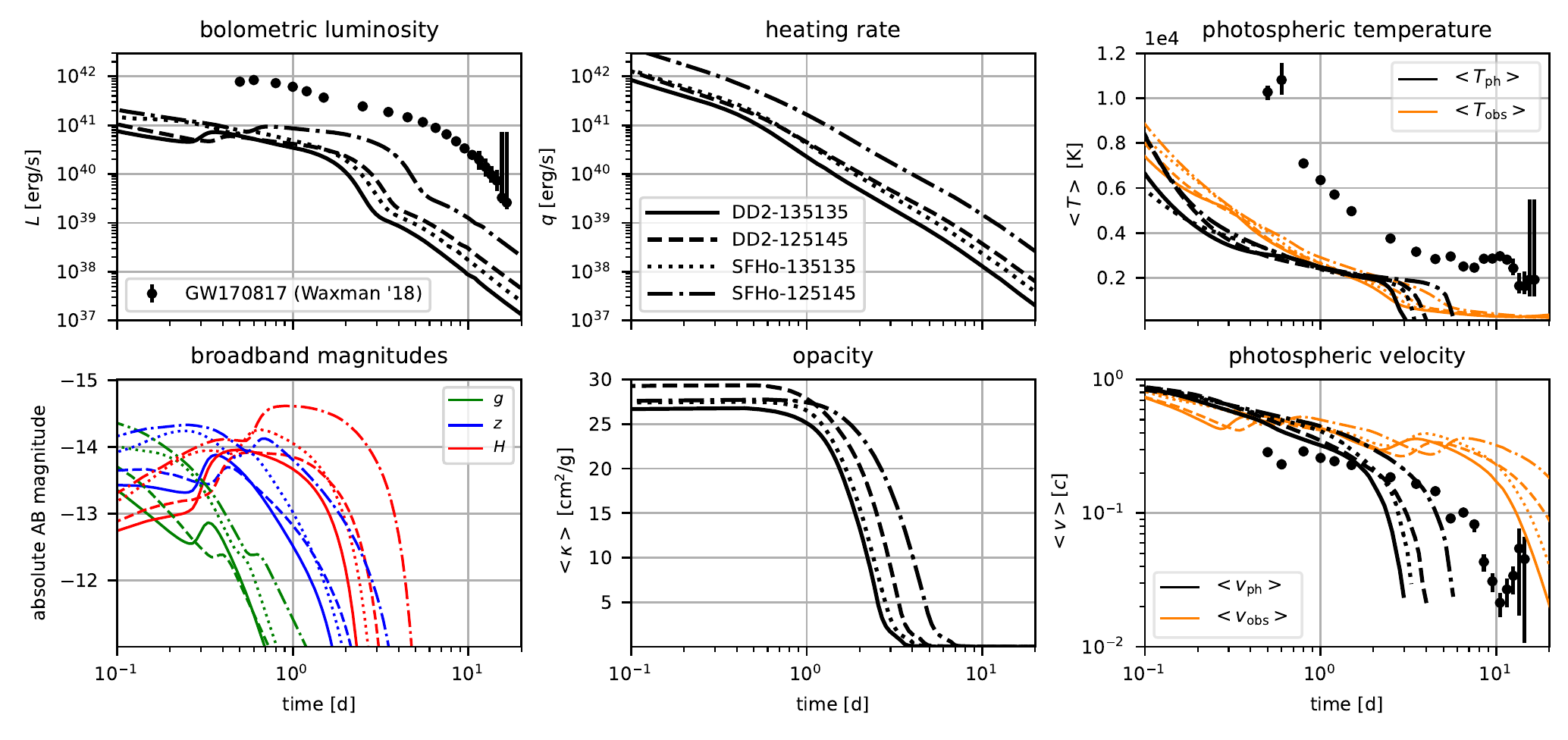}
\caption{Same as Fig.~\ref{fig_tplotpolequ} but comparing the four hydrodynamic models with different nuclear EoSs and binary mass ratios and not distinguishing between polar and equatorial regions, i.e. only spherically averaged or integrated quantities are shown. Note that the top middle panel now shows $q$ instead of $q/M_{\rm ej}$.}
\label{fig_tplots_4mods}
\end{figure*}

We show in Fig.~\ref{fig_tplots_4mods} a similar selection of quantities as in Fig.~\ref{fig_tplotpolequ}, but now for the four models DD2-135135, DD2-125145, SFHo-135135, SFHo-125145 and not distinguishing between polar and equatorial regions in the computation of the corresponding quantities. As shown in Part I, the abundance pattern of nucleosynthesis yields differs only by moderate amounts between these models and, hence, the initial opacities (cf. bottom middle panel of Fig.~\ref{fig_tplots_4mods}) as well as the specific heating rates (see Fig.~8 in Part I) are rather similar for all models. The observed differences in the kilonova emission must therefore be connected to variations of the ejecta masses and velocities. Consistent with previous studies \citep[e.g.][]{Bauswein2013, Hotokezaka2013b, Radice2018b}, the ejecta mass and average velocity increase for the softer SFHo EoS compared to the DD2 EoS, and the ejecta mass increases as well when reducing the mass ratio of the stellar binary. Indeed, the bolometric luminosities of the four models are roughly consistent with the order corresponding to the ejecta masses, at least in the late, optically thin phase of evolution, during which the luminosity is essentially given by the thermal heating power, 
\begin{align}\label{eq:Lthinscaling}
  L\approx L_{\rm thin}\equiv\langle \Qheat\rangle M_{\rm ej}= q \, .
\end{align}
Deviations from an exactly linear proportionality between $L_{\rm thin}$ and $M_{\rm ej}$ at a given time $t$ are mainly connected to the thermalization efficiency, $\fth$, which grows with $M_{\rm ej}$ and decreases with the ejecta velocity, $\langle v\rangle$ (cf. Fig.~\ref{fig_ftherm}). In contrast to the behavior at late times, the luminosity differences right around the peak epochs are significantly smaller. The reason is that more massive ejecta take longer to become optically thin than less massive ejecta, namely roughly until 
\begin{align}\label{eq:tpeakscaling}
  t_{\rm peak} \propto M_{\rm ej}^{\frac{1}{2}} \langle v\rangle^{-\frac{1}{2}} \langle \kappa\rangle^{\frac{1}{2}} 
\end{align}
as estimated based on the simplified case of a uniform distribution of density and opacity \citep[e.g.][]{Arnett1982, Metzger2010c, Grossman2014}. Hence, since the radioactive heating rate declines approximately as
\begin{align}\label{eq:qscaling}
  \langle \Qheat\rangle\approx \langle \fth\rangle \times Q_0 \times \left(\frac{t}{1\,\mathrm{d}}\right)^{-\alpha}
\end{align}
(where $Q_0\approx 2\times 10^{10}\,$erg\,g$^{-1}$\,s$^{-1}$ and $\alpha\approx 1.3$ reproduce well the total radioactive energy release rate, $Q+\Qneu$, resulting from our nucleosynthesis calculations; see Part I\footnote{Note that the analytic fit presented in Part I with a scaling factor of  $1\times 10^{10}\,$erg\,g$^{-1}$\,s$^{-1}$ was for the heating rate without neutrino contributions, $Q$, whereas here we consider $Q+\Qneu$.}), the luminosity at peak epoch scales only weakly with the ejecta mass \citep[e.g.][]{Grossman2014, Kasen2017a, Wollaeger2018a}:
\begin{align}\label{eq:Lpeakscaling}
  L_{\rm peak}\approx & \quad L_{\rm thin}(t_{\rm peak}) \nonumber \\
  \propto & \quad Q_0 \langle\fth\rangle_{\rm peak} M_{\rm ej}^{0.35} \langle v\rangle^{0.65} \langle \kappa\rangle^{-0.65} \, .
\end{align}
Assuming that the implicit dependence of the thermalization efficiency at peak, $\langle\fth\rangle_{\rm peak}\equiv \langle\fth\rangle(t_{\rm peak})$, on $M_{\rm ej}$ and $\langle v\rangle$ is subdominant\footnote{We stress that $\langle\fth\rangle(t_{\rm peak})$ varies much less between individual models than $\langle\fth\rangle(t)$ at some fixed time, $t$, as can be seen from Fig.~\ref{fig_ftherm} using the peak times provided in Table~\ref{tab:prop}.}, the behavior of $t_{\rm peak}$ and $L_{\rm peak}$ in our models (cf. Table~\ref{tab:prop}) can be understood reasonably well by means of  Eqs.~(\ref{eq:tpeakscaling}) and (\ref{eq:Lpeakscaling}), respectively.

In Table~\ref{tab:prop} we also provide the peak photospheric temperatures, $\langle T_{\rm ph}\rangle_{\rm peak}$, and find that they all lie rather close to each other within $2.2\ldots 2.7\times 10^3\,$K, i.e. neither very sensitive to the nuclear EoS nor to the binary mass ratio. The weak dependence on $M_{\rm ej}$ and $\langle v\rangle$, is in agreement with expectations from an analytical estimate of the peak photospheric temperature, which can be derived using the Stefan-Boltzmann law under the assumption that the ejecta emit black-body radiation from their surface at about $r\approx \langle v\rangle t$ \citep[e.g.][]{Grossman2014}:
\begin{align}\label{eq:Tpeakscaling}
  \langle T_{\rm ph}\rangle_{\rm peak}\approx & \quad \left[\frac{L_{\rm peak}}{\sigma_{\rm SB}4\pi (\langle v\rangle t_{\rm peak})^2} \right]^{\frac{1}{4}}\nonumber \\
  \propto &\quad Q_0^{0.25} \langle\fth\rangle_{\rm peak}^{0.25} M_{\rm ej}^{-0.16} \langle v\rangle^{-0.09} \langle \kappa\rangle^{-0.41} 
\end{align}
(where $\sigma_{\rm SB}$ is the Stefan-Boltzmann constant). Measured in terms of the power by which they determine $\langle T_{\rm ph}\rangle_{\rm peak}$ in this simplified analytical estimate, the mass and velocity of the ejecta are less important than the opacity and heating rate. Not surprisingly, considering the rather similar values of $\langle T_{\rm ph}\rangle_{\rm peak}$, all models peak in the same ($H$) frequency band.

In summary, in our kilonova models the luminosity (spectral temperature) is rather sensitive (insensitive) to variations of the nuclear EoS and the binary mass ratio. Since the composition is fairly independent of the EoS and mass ratio, the bolometric luminosity is therefore mainly determined by the ejecta mass and velocity. We note that the simplified grey opacities used here are not suited for an in-depth analysis of spectral features. More sophisticated calculations using opacities based on atomic lines may reveal sensitivities that are obscured by our scheme.

\subsubsection{Ignoring neutrino interactions}\label{sec:ignor-neutr-inter}

\begin{figure*}
\includegraphics[width=0.99\textwidth]{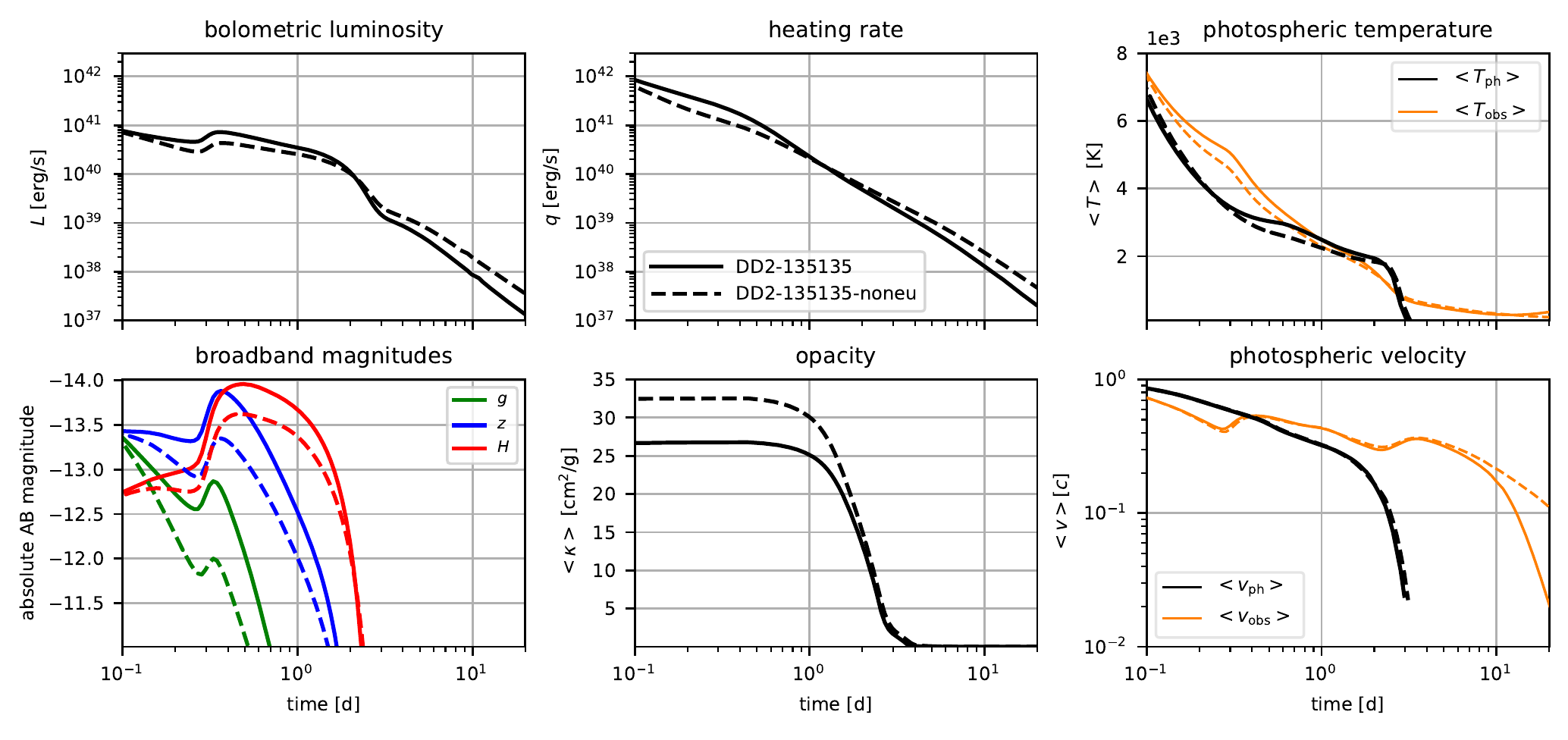}
\caption{Same as Fig.~\ref{fig_tplots_4mods} but comparing the fiducial model DD2-135135 with its counterpart DD2-135135-noneu, in which neutrino reactions have been neglected after the plunge.}
\label{fig_tplots_noneu}
\end{figure*}

The impact of neutrino interactions on the composition and the kilonova of dynamical ejecta (as well as other ejecta components) represents one of the most uncertain and challenging aspects in theoretical multi-messenger modeling of NS mergers given the difficulties of solving the Boltzmann equation for neutrinos \citep[e.g.][]{Wanajo2014a, Goriely2015, Martin2018a, Radice2018b, Foucart2018b, Sumiyoshi2021a}.
Until a few years ago most merger simulations therefore either completely neglected neutrino interactions or implemented them in a way that neglects net neutrino absorption and/or the advection and equilibration of trapped neutrinos. The advection of trapped neutrinos can be accounted for without significant efforts and even without computing any neutrino interaction rates -- basically by replacing the electron fraction with the total lepton fraction in optically thick regions \citep[e.g.][]{Goriely2015, Ardevol-Pulpillo2019a}. However, the main challenge lies in the reliable description of the emission and absorption rates in the semi-transparent regions surrounding the hot merger remnant once it is formed. In order to quantify the impact of (not) including any neutrino interactions after the first touch of the two stars, we introduced in Part I a variation of model DD2-135135 (called DD2-135135-noneu here), in which $Y_e$ is held constant after the point when the two stars plunge into each other. In this model the abundances of elements with mass numbers $90\la A\la 140$ are found to be reduced, while the mass fractions of lanthanides and heavier elements are enhanced by a factor of $2-3$. Moreover, the pole-to-equator composition gradient basically disappeared and high values of $\XLA\ga 0.1$ are now found at all latitudes (see fifth row of Fig.~\ref{fig_cont}). For the heating rate the lack of low-mass elements causes a reduction at early times ($0.1\,\mathrm{d}\la t\la 1\,\mathrm{d}$), while the excess of heavy elements produces an increase at late times ($1\,\mathrm{d}\la t\la 100\,\mathrm{d}$) compared to the full neutrino model (cf. Sect.~3.2 and Fig.~8 of Part I).

Figure~\ref{fig_tplots_noneu} provides insight about the extent to which the kilonova light curve and related quantities are affected by neglecting neutrino interactions during (most of) the merger process. Initially, when the temperatures are above the recombination temperatures, the enhancement factor of $\sim 2$ in the lanthanide fraction translates into an opacity increase of $\approx 20\,\%$. The weaker radioactive heating under more opaque conditions at $t< 1\,$d explains why the emission peak is fainter by about $30-50\,\%$ and is delayed by $\sim 10\,\%$ (cf. top left panel of Fig.~\ref{fig_tplots_noneu} and Table~\ref{tab:prop}). At late times, when opacity effects have ceased and heavier elements dominate the heating rate, the situation is reversed because of the higher abundance of heavier elements in the ``noneu'' model. Considering broadband light curves, the kilonova peak remains in the red regime of the near-IR band, represented here by the $H$-band, while its magnitude is reduced by about 0.3\,mag. The photospheric temperatures near peak epoch are reduced, though only mildly.

\subsubsection{Sensitivity to $Y_e$}\label{sec:sensitivity-y_e}

\begin{figure*}
\includegraphics[width=0.99\textwidth]{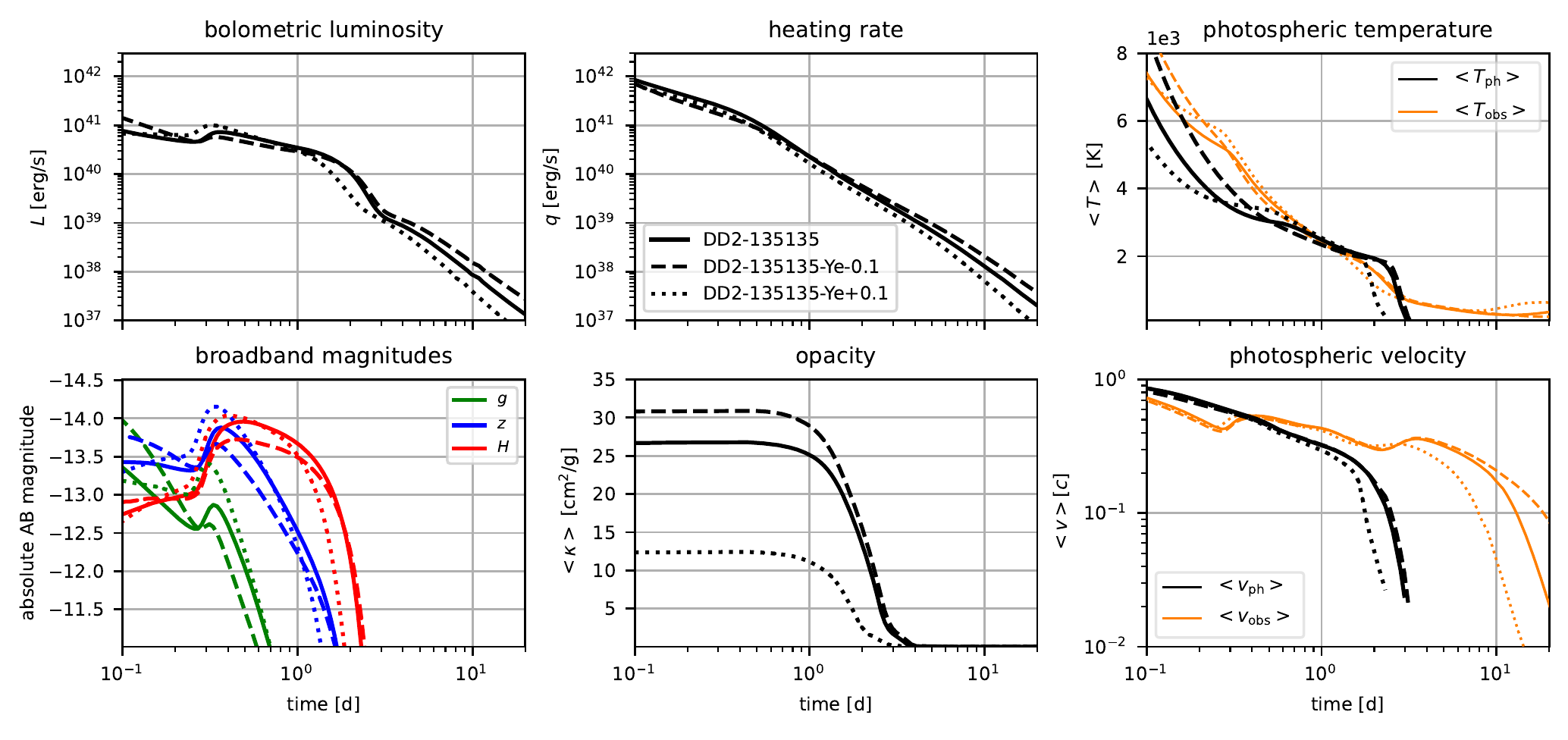}
\caption{Same as Fig.~\ref{fig_tplots_4mods} but comparing the fiducial model DD2-135135 with its counterparts DD2-135135-Ye-0.1 and DD2-135135-Ye+0.1, in which the electron fraction, $Y_e$, was shifted by $-0.1$ ($+0.1$) for all trajectories.}
\label{fig_tplots_ye}
\end{figure*}

An additional sensitivity test to bracket the uncertainties in the treatment of weak interactions can be conducted by manually varying the electron fraction at the onset of the nuclear network calculations, $Y_e(\rho_{\rm net})$. To this end, we set up two models that are identical to the fiducial model but where $Y_e$ of all trajectories is increased and reduced, respectively, by 0.1 (with a lower limit of 0.005 imposed). The nucleosynthesis yields of these models have already been discussed in Part I, see Sect.~3.3 and Fig.~11 therein.

In Fig.~\ref{fig_tplots_ye} we compare both models with the fiducial model. Artificially reducing $Y_e$ by 0.1 yields results that are very similar to the case of completely neglecting neutrino interactions (see previous section and cf. Fig.~\ref{fig_tplots_noneu}), namely a delayed and fainter peak. This is not surprising considering that the average electron fractions of the corresponding models are comparable ($\langle Y_e\rangle\approx 0.17$ and 0.13, respectively; cf. Table~\ref{tab:prop}). In the opposite case of increasing $Y_e$ by 0.1, the mass fraction of lanthanides and the opacity are reduced roughly by factors of 4 and 2, respectively. The lower opacities result in a significantly shorter duration of the kilonova and a slightly more luminous peak. The luminosity enhancement due to the reduced opacity is, however, partially compensated by the weaker radioactive heating rates in the $Y_e$+0.1 model. At $t=1\,$d, for instance, the heating rate is about $q/(10^{40}\,\mathrm{erg\,s}^{-1})=1.66$ compared to 2.32 in the reference model. The reduced opacities also shift the spectral peak to the blue end of the near-IR domain, as can be seen by the enhanced $z$-band magnitudes in the lower left panel of Fig.~\ref{fig_tplots_ye}. Yet, despite the considerable shift of $Y_e$ and resulting reduction of the lanthanide fraction down to $\langle \XLA\rangle\approx 0.02$, the emission in the visible bands is still subdominant compared to that in the near-IR domain, i.e. the transient would classify as a red kilonova.

Summarizing the results of Sects.~\ref{sec:ignor-neutr-inter} and~\ref{sec:sensitivity-y_e}: Differences in the treatment of neutrino effects can induce variations in brightness of up to a factor of two and in duration of $\sim 20\,\%$, i.e. an amount comparable to that resulting from changing the binary mass ratios or the nuclear EoSs.

\section{Discussion}\label{sec:discussion}

\subsection{Role of dynamical ejecta in AT2017gfo/GW170817}\label{sec:role-dynam-ejecta}

A comprehensive understanding of the origin of each component in the kilonova accompanying GW170817, AT2017gfo, is still elusive. Although the lack of the secular post-merger ejecta in our study and the small set of considered EoSs and NS merger configurations defies a conclusive comparison, we can nevertheless briefly speculate whether our models, which in contrast to most previous models are based on consistent nucleosynthesis trajectories from hydrodynamical simulations, would, or not, support the possibility that the dynamical ejecta \emph{alone} are responsible for either of the two, red or the blue, kilonova components. The observed bolometric luminosity of AT2017gfo, as constructed by \citet{Waxman2018a}, is shown by filled circles in Fig.~\ref{fig_tplots_4mods} as well as estimates of the spectral temperature and photospheric velocity, which can be compared directly to $\langle T_{\rm obs}\rangle$ and $\langle v_{\rm obs}\rangle$ (orange lines in the right panels of Fig.~\ref{fig_tplots_4mods}). As can be seen, none of our models appears to be compatible with either component of AT2017gfo.

The early, blue component of AT2017gfo, i.e. the emission observed until $t\sim 1-1.5\,$d \citep[e.g.][]{Nicholl2017m}, exhibited a bolometric luminosity of almost one order of magnitude higher than the surface-averaged values predicted by our models, and its spectral temperatures were roughly twice as high as in our models. The combination of both discrepancies, too faint luminosities and too red colors, could hardly be resolved by just increasing the ejecta mass, because more massive ejecta would rather reduce the peak temperatures, as suggested by Eq.~(\ref{eq:Tpeakscaling}) neglecting uncertainties connected to the thermalization efficiency.
  The comparison suggests that, in agreement with previous studies based on more simplified one-zone models \citep{Villar2017a}, significantly lower lanthanide mass fractions, and corresponding opacities, may be required to reproduce the early component. Even though the average electron fraction is raised significantly as a result of a sophisticated inclusion of neutrinos in our models (e.g. from $\langle Y_e\rangle=0.13$ to 0.27 for model DD2-135135; cf. Table~\ref{tab:prop}), the amount of lanthanides still seems to be too high to produce an optical transient like the one in AT2017gfo. However, our set of investigated models is rather small, and more exhaustive explorations using other EoSs and NS binary configurations compatible with GW170817 are needed to clarify whether cases exist in which the lanthanide mass fraction becomes significantly smaller than observed for our models.

The relatively red spectral colors seen in the kilonova signals of our models suggests a better agreement of our models with the red component of AT2017gfo \citep[e.g.][]{Chornock2017a}\footnote{We remark that post-merger disk outflows are often considered more suitable to explain the red component \citep[e.g.][]{Kasen2017a}, because they are typically more massive than the dynamical ejecta while carrying comparable lanthanide fractions \citep[e.g.][]{Just2015a, Wu2016a, Siegel2017b, Just2021a}. See, e.g., \citet{Tanaka2017t, Perego2017a, Kawaguchi2018a} for the discussion of alternative scenarios.}, especially keeping in mind that (cf. Fig.~\ref{fig_angdep}) isotropic luminosities observed from polar viewing angles can be enhanced by some factor compared to the surface-averaged isotropic luminosities plotted in Fig.~\ref{fig_tplots_4mods}. However, the ejected mass would presumably need to be significantly larger (by a factor of a few) than the $0.009\,M_\odot$ of our most massive model in order for the peak to occur sufficiently late and luminous (cp. Fig.~\ref{fig_tplots_4mods} and Eqs.~(\ref{eq:tpeakscaling}) and (\ref{eq:Lpeakscaling})). This could possibly be achieved by very asymmetric cases and a soft equation of state \citep[e.g.][]{Bauswein2013, Sekiguchi2016a, Radice2018b}. Moreover, a crucial role may be played by the uncertainties associated with the thermalization efficiency, $\fth$ \citep{Barnes2016a, Kasen2019a, Wu2019c}. The fact that $\fth$ increases with the ejecta mass could improve the prospects in this regard. 

In any case, our analysis remains incomplete and inconclusive, because we only consider a small set of EoSs and binary configurations and we only include a single ejecta component. A more realistic model of AT2017gfo would need to account for genuine multi-component effects, such as photon transport from one ejecta component to another \citep[e.g.][]{Kawaguchi2018a}, or photon blocking of one component by another \citep[often called ``lanthanide curtain''; see, e.g.,][]{Kasen2015, Korobkin2021e}.

\subsection{Importance of consistent ejecta sampling}\label{sec:import-cons-ejecta}

\begin{figure*}
\includegraphics[width=0.22\textwidth]{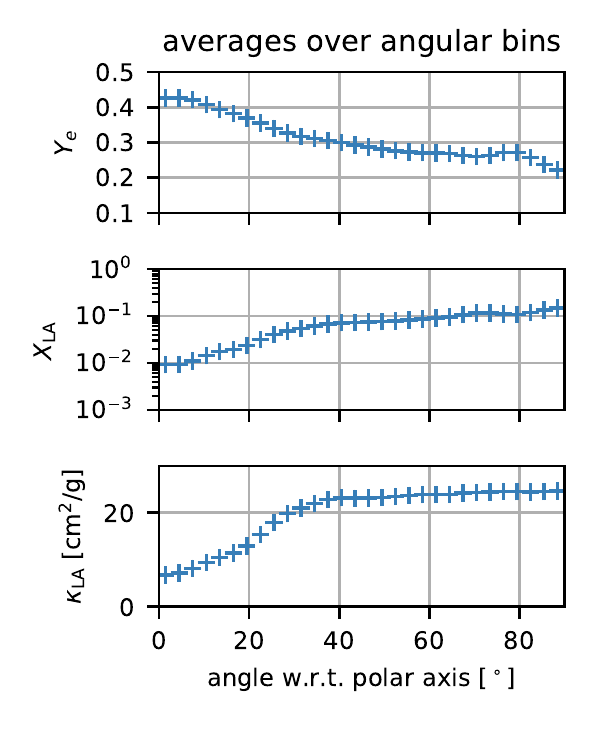}
\includegraphics[width=0.77\textwidth]{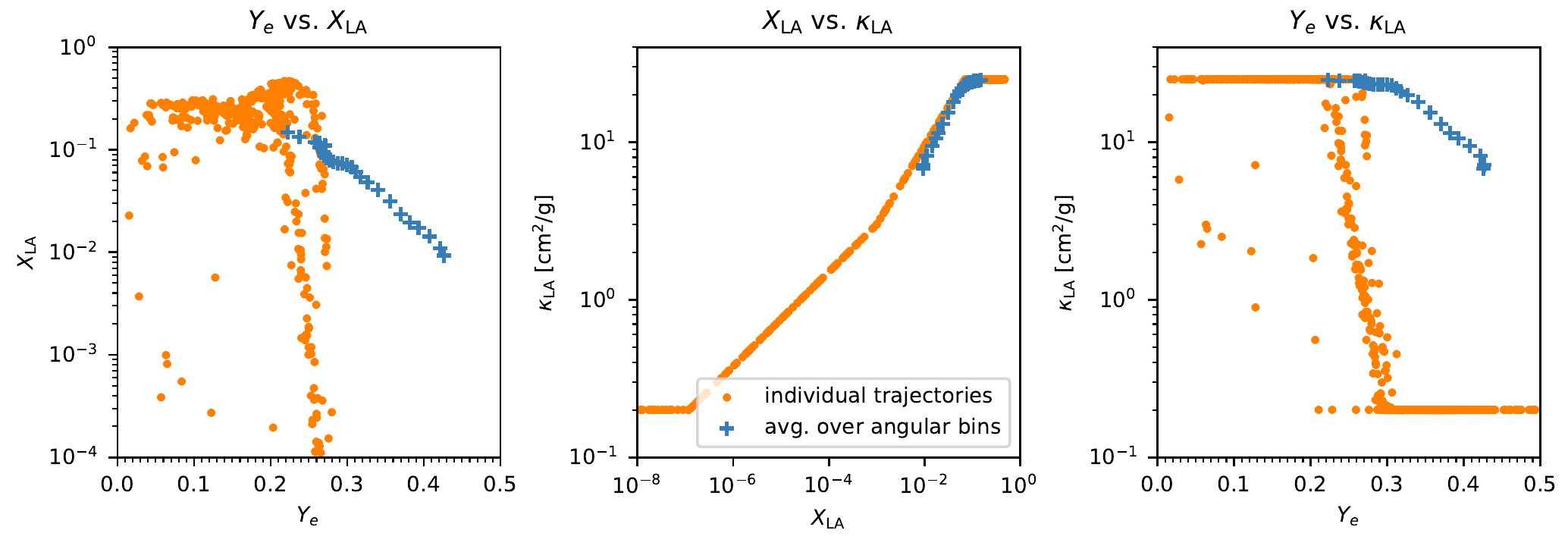}
\caption{Example illustrating the discrepancy between correlations obtained for individual trajectories and correlations between averages over finite domains. All plotted data is taken from model DD2-135135. \emph{Three left plots:} Mass-averages of the electron fraction, $Y_e$, lanthanide mass fraction, $\XLA$, and (temperature-independent part of the) opacity, $\kapla$, over the volumes defined by 30 equidistant polar-angle bins dividing the northern hemisphere. \emph{Three right plots:} Pair-wise correlation plots between the same data as in the left plots. For comparison, orange dots denote the corresponding data measured for individual outflow trajectories. A sharp cut-off of $\XLA$ around $Y_e\sim 0.28$ is only visible for single-trajectory data, while much higer values of $\XLA(Y_e)$, and therefore of $\kapla(Y_e)$, can be reached when considering collections of trajectories in finite regions.}
\label{fig_opacavg}
\end{figure*}

\begin{figure}
\includegraphics[width=0.48\textwidth]{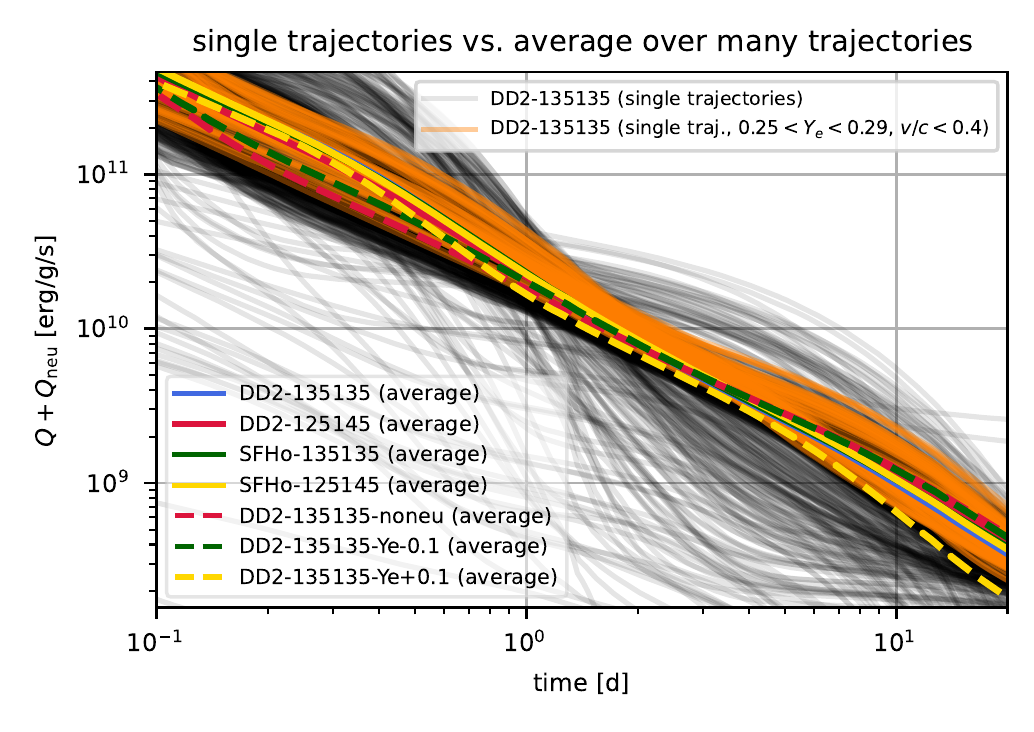}
\caption{Comparison of total radioactive energy-release rates, $Q+\Qneu$, along individual outflow trajectories of model DD2-135135 and the corresponding rates mass-averaged over all trajectories of a given model. Orange lines denote the subset of particles of model DD2-135135 with $Y_e$ close to the average value of $\langle Y_e\rangle=0.27$ and with velocities $v/c<0.4$. Values of $Q+\Qneu$ differ more strongly between individual trajectories than between ensemble averages of different models, even for models with rather diverse values of $0.13<\langle Y_e \rangle<0.37$.}
\label{fig_heatavg}
\end{figure}

An often employed approximation in modeling ejecta and kilonovae from NS mergers is to assume representative bulk values of the electron fraction, expansion velocity, and entropy for the entire ejecta, or for sub-domains within the ejecta, and to derive from those the composition, heating rate, and opacity using a single nucleosynthesis calculation. Obviously, such single-trajectory models carry some uncertainty, because in reality the ejecta exhibit a broad spectrum of hydrodynamic and thermodynamic properties, which may result in large variations of heavy-element yields among different fluid elements in the ejecta. The yields obtained by averaging over a collection of nucleosynthesis trajectories may predict results that cannot be accounted for by a single nucleosynthesis calculation along a trajectory of averaged fluid properties. Our method of detailed ejecta sampling by post-processing individual outflow trajectories using consistent hydrodynamic data allows to test the uncertainty of single-trajectory models by comparing the average opacities and heating rates predicted by either approach.

We first consider the opacity. It is well known \citep[e.g.][]{Ruffert1997, Hoffman1997, Lippuner2015a, Lemaitre2021m} that the production of lanthanides along a typical outflow trajectory from NS mergers becomes inefficient as soon as the electron fraction exceeds a threshold value, $Y_e^{\rm crit}$, which, for typical merger outflows and mainly depending on the entropy per baryon, lies within $Y_e^{\rm crit}=0.25-0.3$. As can be seen by the dots in the middle panel of Fig.~\ref{fig_opacavg}, this behavior is reproduced well by nucleosynthesis yields for individual ejecta trajectories of our model DD2-135135: All trajectories with $Y_e\ga 0.28$ exhibit extremely low lanthanide fractions of $\XLA<10^{-4}$. This step-like behavior is often used to characterize also the bulk ejecta or sub-regions of the ejecta by assigning very low values of $\XLA$ to regions with average $Y_e>Y_e^{\rm crit}$. However, if instead of individual trajectories we consider finite sub-domains of the ejecta, for example obtained by dividing the northern hemisphere into 30 angular bins (see blue crosses in Fig.~\ref{fig_opacavg}), we receive quite a different relation between $Y_e$ and $\XLA$. Due to mixing of low-$\XLA$ fluid elements with high-$\XLA$ fluid elements, the sharp cut-off is replaced by a much more shallow decline, such that regions with average $Y_e>0.4$ can still contain a high fraction of lanthanides $\XLA>0.01$ and posess a correspondingly high opacity. If, for instance, we had estimated the opacity of angular bins based on the $\kapla(Y_e)$ behavior of individual trajectories, e.g. by using $\kappa_{\rm LA}(Y_e>0.28)=0.2$ and $\kappa_{\rm LA}(Y_e<0.28)=30\,$cm$^{-2}$g$^{-1}$, the average opacity of the polar ($\theta<\pi/4$) ejecta would be $0.2\,\,$cm$^{-2}$g$^{-1}$, i.e. grossly underestimated compared to the value of $\approx 6.2\,\,$cm$^{2}$g$^{-1}$ that would result if we had used the same $\kapla(Y_e)$ function for individual particles instead. Hence, approximating $\XLA$ or $\kapla$ based on the average $Y_e$ of an ensemble of trajectories can lead to a significant underestimation of $\XLA$ or $\kapla$, which tends to be more serious for a higher degree of inhomogeneity of the ejecta. While here we consider only the dynamical ejecta, this aspect is likely to be relevant also for turbulence-driven secular ejecta, which typically exhibit a broad spectrum of $Y_e$ \citep[e.g.][]{Fernandez2013b, Siegel2017b, Just2021a}, while it might be less relevant for neutrino-driven winds, which often have a rather smooth structure \citep[e.g.][]{Perego2014a, Fujibayashi2018a}.

Next we take a look at the radioactive heating rates. As is already clear from the analytic scaling laws for $L_{\rm thin}$ and $L_{\rm peak}$ (cf. Eqs.~(\ref{eq:Lthinscaling}) and~(\ref{eq:Lpeakscaling}), respectively) a reliable prediction of the heating rate is mandatory for any kilonova model to properly infer the ejecta mass from observations during the peak or the optically thin phase \citep[e.g.][]{Wanajo2018a, Rosswog2018a, Wu2019c, Barnes2021c}. For simplicity, we now ignore the additional complexity connected to the thermalization efficiency and assume $\fth=1$. For the heating rates the analysis is not as straightforward as for the opacities, where the strong non-linearity of $\XLA(Y_e)$ represents the main culprit, because the heating rates are determined by decay chains involving a large number of isotopes, of which the abundances are quite sensitive also to the entropy and expansion timescale. Nevertheless, we can get a basic idea of the uncertainty of single-trajectory models by comparing the heating rates for individual trajectories with the average heating rates. To that end, in Fig.~\ref{fig_heatavg} we plot, apart from the global heating rates for each model, the heating rates of all particles of model DD2-135135 using black, slightly transparent lines, while orange lines are used to denote only the subset of these particles that have $Y_e$ close to the average value of $\langle Y_e\rangle\approx 0.27$ and velocities of $v/c<0.4$. Even for this subset of presumably representative trajectories (in the sense that $Y_e$ and $v$ are in the ballpark of expected bulk values), the range of variation of the heating rates is substantial and amounts to factors of 2-4 during the entire period of $0.1\,$d$<t<20\,$d that is relevant for the kilonova emission. On the other hand, the substantially smaller model-by-model variations of the average heating rates (thick lines in Fig.~\ref{fig_heatavg}) indicate that the average heating rate of ensembles of trajectories are subject to a much reduced level of randomness than the heating rate of individual particles. This finding is particularly remarkable, because we even consider models with artificially changed values of $Y_e$ that span a large range of values for $\langle Y_e\rangle$ between 0.13 and 0.37 (cf. Table~\ref{tab:prop}).

From the above discussions we conclude that predictions for the nucleosynthesis yields and the corresponding kilonova signal that are based on one-zone models or single-trajectory modeling may carry substantial systematic uncertainties in cases where the thermodynamic properties are not homogeneous throughout the ejecta but given by a broad distribution.

\subsection{Comparison with previous studies}\label{sec:comp-with-prev}

\begin{figure}
\includegraphics[width=0.499\textwidth]{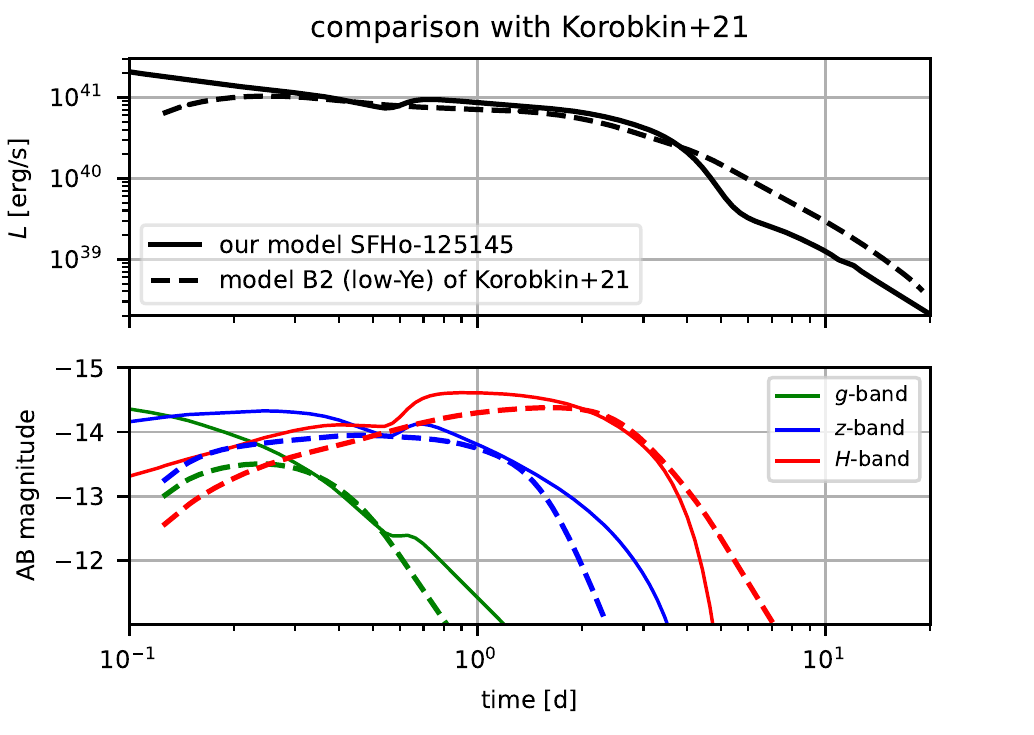}
\caption{Exemplary comparison of the surface-integrated bolometric luminosity \emph{(top panel)} and the AB magnitudes in selected bands \emph{(bottom panel)} of model B2 (low-$Y_e$ version) of \citet{Korobkin2021e} with our model SFHo-125145. The good agreement with \citet{Korobkin2021e}, who employ a Monte-Carlo radiative-transfer scheme with atomic-physics based opacities supports the validity of our results obtained with an approximate radiative transfer solver with parametrized opacity treatment.}
\label{fig_korobkin}
\end{figure}

\begin{figure}
\includegraphics[width=0.48\textwidth]{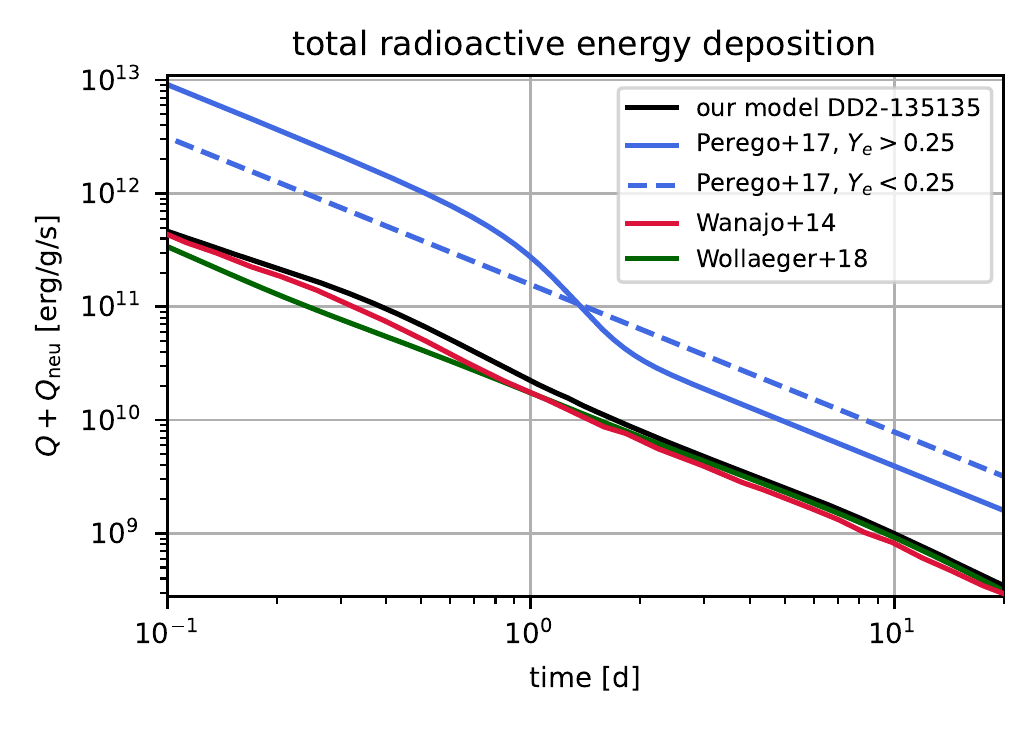}
\caption{Comparison of mass-averaged total radioactive energy-release rates resulting from our nucleosynthesis calculations for model DD2-135135 with heating rates employed in other kilonova studies (not accounting for thermalization efficiencies). The red curve (extracted from Fig.~5 of \citealp{Wanajo2014a}) denotes the heating rate obtained from a nucleosynthesis analysis of a neutrino-hydrodynamical merger simulation and was employed for dynamical ejecta in, e.g., \citet{Kawaguchi2018a, Kawaguchi2020h}. The green line (extracted from Fig.~4 of \citealp{Wollaeger2018a}) is based on a network calculation along an exemplary outflow trajectory of model B of \citet{Rosswog2014} and was employed for dynamical ejecta in, e.g., \citet{Wollaeger2018a, Korobkin2021e, Wollaeger2021a}. Blue lines (cf. Eq.~2 and case BF of \citealp{Perego2017a}) represent the $Y_e$-dependent prescription employed by \citet{Perego2017a, Radice2018b}, which is not based on a nuclear network calculation but deduced from a light curve fit to AT2017gfo.}
\label{fig_peregoheat}
\end{figure}

We will now briefly compare some of our results with previous studies. The level of sophistication of kilonova models is growing quickly, and many studies employ their own ways of dealing with the large number of required physics ingredients. Moreover, given the large parameter space it is difficult to find studies that discuss models with a match of the basic characteristic parameters (geometry, mass, velocity, composition, etc.), which would be required for a meaningful quantitative comparison. Hence, we will only compare a few elementary features with selected studies.

While the number of kilonova studies directly based on the outputs from hydrodynamical simulations is still very small, quite a few studies exist by now that investigate multi-dimensional kilonovae based on manually constructed ejecta configurations with homogeneous distributions of the heating rate and opacity \citep[e.g.][]{Wollaeger2018a, Barbieri2019a, Bulla2019a, Darbha2020b, Korobkin2021e, Kawaguchi2020h, Heinzel2021e}. \citet{Darbha2020b} studied the emission characteristics as function of the observer angle based on parametrized ellipsoids, tori, and conical sections with a grey and constant opacity and fitted the results with an analytic function for the ratio $L_{\rm iso}(0^\circ)/L_{\rm iso}(90^\circ)$ in dependence of the projected surface area of the emitting object. For an ellipsoid with a semi-major axis ratio of $\sim 1.5-2$, which to first order resembles the shape of the photospheres that we observe in our models (cf. Fig.~\ref{fig_cont}), they predict values of $L_{\rm iso}(0^\circ)/L_{\rm iso}(90^\circ)\approx 1.5-3$ that are in good agreement with our results (see Fig.~\ref{fig_angdep}).

Various types of parametrized, axisymmetric ejecta configurations have been employed in a series of papers by the Los Alamos group \citep[e.g.]{Wollaeger2018a, Korobkin2021e, Wollaeger2021a} using atomic-physics based opacities \citep{Fontes2020d} and two different choices for the composition \citep{Even2020a}. The ellipsoidal geometry of photospheres found here for the dynamical ejecta is roughly reproduced by the biconcave (``B'') configuration introduced by \citet{Korobkin2021e}, which however was dropped again in the set of models by \citet{Wollaeger2021a}. Out of the single-component models considered by \citet{Korobkin2021e}, their model B2, with an average velocity of $0.2\,c$ and a mass of 0.01$\,M_\odot$, posesses parameters with the closest agreement to one of our models, namely to SFHo-125145. A caveat for our comparison is, however, that their ``low-$Y_e$'' case exhibits a lower lanthanide fraction ($\XLA=0.032$; see \citealp{Korobkin2021e} and \citealp{Even2020a}) compared to our model (0.114). In Fig.~\ref{fig_korobkin} we compare\footnote{The data for model B2 was downloaded from https://ccsweb.lanl.gov/astro/transient/transients\_astro.html.} the angle-averaged bolometric light curves and magnitudes for both models and find rather good agreement, both quantitatively and qualitatively regarding the property that the emission peaks in the same near-IR frequency regime.

A similar approach as in the aforementioned works was pursued in \citet{Kawaguchi2018a, Kawaguchi2020h}, however, almost exclusively considering multi-component (dynamical plus secular) ejecta models. Broadband magnitudes for two single-component models of dynamical ejecta, with masses of 0.003 and 0.01\,$M_\odot$ and average velocity of 0.25$\,c$ are presented in Appendix A of \citet{Kawaguchi2020h}. Despite the fact that their setup agrees fairly well with ours -- the main difference is probably the more toroidally shaped ejecta distribution in their case -- we find only modest agreement. Independent of the observer angle their kilonova is brighter by about 1\,mag in the near-IR bands and, at early times and polar viewing angles, they obtain a considerably stronger optical component than us. The origin of these differences remains unclear but could be connected to the more advanced set of opacities \citep[which are based on][]{Tanaka2020a} employed in these studies and to the different ejecta geometry.

Among the few available kilonova models that adopt ejecta properties directly from hydrodynamical simulations are \citet{Kawaguchi2021m} and \citet{Radice2018b}. A direct comparison with the light curves of \citet{Kawaguchi2021m} is not possible, however, because they additionally follow the long-term evolution of the NS-disk remnant, which produces massive viscously driven outflows that dominate the dynamical ejecta. Yet, a noteworthy difference to the models in \citet{Kawaguchi2021m} is the prolate geometry of their total ejecta, which is in contrast to the rather oblate structures that we find for just the dynamical ejecta at peak epoch.

\citet{Radice2018b} reported kilonova light curves that are obtained by a superposition of quasi-spherically symmetric models adopting the formalism by \citet{Grossman2014, Martin2015, Perego2017a}. \citet{Radice2018b} used data provided by binary NS merger simulations for the emission from the dynamical ejecta, while they employed a parametric wind model to estimate the contribution from neutrino- plus viscously-driven post-merger ejecta. Although no light curves just for the dynamical ejecta components are shown in \citet{Radice2018b}, one can indirectly infer from their results that the dynamical ejecta produce significant emission in the $g$ and $z$ bands, with absolute AB magnitudes of about -15\,mag (corresponding to apparent magnitudes of $\approx 18\,$mag when observed at a distance of $40\,$Mpc) for dynamical ejecta masses of $\approx 0.003\,M_\odot$. Our models, which seem to broadly agree in the mass- and $Y_e$-distributions of the ejecta, instead produce significantly fainter emission which peaks in bands with lower frequency. One reason for the higher luminosities may be that \citet{Radice2018b} assigned the opacities as function of average $Y_e$ values along angular bins (see Sect.~\ref{sec:import-cons-ejecta} for a discussion). Another, more likely explanation may come from the fact that \citet{Radice2018b} employed the extremely high heating rates by \citet{Perego2017a}. These rates, which are shown in Fig.~\ref{fig_peregoheat}, were not adopted from nucleosynthesis calculations but were calibrated such that the semi-analytic kilonova models by \citet{Perego2017a} reproduce the observed kilonova AT2017gfo\footnote{A similar approach involving even higher heating rates was recently adopted by \citet{Breschi2021t}.}. Switching to such a high heating rate enhances the peak luminosities substantially, because the heating rate enters $L_{\rm peak}$ linearly, whereas the ejecta mass, velocity, and opacity only enter with a lower power (cf. Eq.~(\ref{eq:Lpeakscaling})). Hence, we expect the heating rate to be the main reason for the observed discrepancies between their kilonova results and ours.

Although the high heating rates assumed in \citet{Perego2017a} cannot be excluded entirely, to our knowledge no state-of-the-art network solver currently exists that produces heating rates of that magnitude for the thermodynamic conditions under consideration. A systematic discussion of the sensitivity of the heating rate with respect to nuclear-physics uncertainties is, however, deferred to a future study. We point out that \citet{Barnes2021c} and \citet{Zhu2021a} have recently taken a step in that direction and report large variations between $Q$ profiles with different nuclear physics inputs. However, the range of uncertainty may not be as dramatic as suggested by these studies, because, first, some of their most extreme cases are produced by nuclear models, which are known to be inconsistent with currently known experimental data\footnote{For example, \citet{Barnes2021c} and \citet{Zhu2021a} employ the SLy4 HFB masses, which are known to predict all experimental masses of even-even nuclei with a root-mean-square deviation as large as 5 MeV \citep{Stoitsov2003x}.}, and second, their analysis is based on single outflow trajectories instead of trajectory ensembles, which may artificially enhance the range of scatter as indicated by the comparison discussed in Sect.~\ref{sec:import-cons-ejecta} (Fig.~\ref{fig_heatavg} there).

\section{Summary and conclusions}\label{sec:summary-conclusions}

In this study we applied a new methodical treatment to predict the kilonova emission of dynamical ejecta adopting tracer particles from general relativistic neutrino-hydrodynamics simulations. In contrast to many previous kilonova models using only the bulk properties of the ejecta, the tracer-particle approach samples the local, time-dependent properties of the ejected material as well as the corresponding nucleosynthesis results for the radioactive heating rate, lanthanide mass fraction, and mean baryon number. The kilonova emission is computed by means of a two-moment transport scheme employing the approximate M1 closure and assuming a simplified, heuristic expression of the photon opacity as function of the lanthanide fraction and the temperature. We applied our approach to the dynamical ejecta of four different NS-merger configurations with different nuclear equations of state and different binary mass ratio. Moreover, in order to quantify the sensitivity with regard to uncertainties of weak interaction effects taking place during the dynamical merger phase, we compared with three additional cases with artificially modified electron fractions. Our results are summarized as follows:
\begin{itemize}
\item At times close to peak emission, all models exhibit photospheres that globally resemble ellipsoids with semi-major axis ratio of $1.2-2$. The globally oblate geometry of the photosphere is a result of less efficient matter ejection towards the poles as well as systematically higher electron fractions, and therefore reduced opacities, in the polar compared to equatorial regions. Due to the violent, highly dynamic matter-ejection processes during the merger, the photosphere is not smooth but rather corrugated and characterized by irregular structures that change with time as the photosphere travels deeper into the ejecta.
\item For the four dynamical ejecta models with regular treatment of weak interactions, masses between $0.002-0.009\,M_\odot$, average velocities of $\approx 0.25\,c$, and lanthanide mass fractions of $\approx 0.1$ the ejecta reach optically thin conditions at times between $0.7-1.5$\,d, while during that epoch they produce bolometric luminosites of $\sim 3-7\times 10^{40}\,$erg\,s$^{-1}$, which peak in near-IR frequency bands, and they exhibit photospheric temperatures within $2.2-2.8\times 10^3\,$K. Since the composition is rather similar for the four regular models, the observed differences of peak times and luminosities between these models are a result mainly of the different ejecta masses. A softer equation of state as well as a more asymmetric binary produce slightly more luminous transients with longer durations.

\item The fluxes received by observers increase by a factor of $1.5-3$ when going from equatorial to polar viewing angles, while for different configurations this factor increases with the level of oblateness of the ejecta. The spectra differ only marginally for different observer angles.

\item Neglecting all neutrino interactions after both NSs first touched for a symmetric model with DD2 EoS reduces the average electron fraction from 0.27 to 0.13, increases the lanthanide fractions by a factor of about 2, and moves radioactive heating power effectively from times $t<1\,$d to $t> 1\,$d. The main consequences for the kilonova are an extended peak (by $\sim 15\,\%$) and a luminosity reduction (by $\sim 40\,\%$), while the spectrum is only mildly red-shifted. These changes are comparable to the changes obtained when reducing $Y_e$ by 0.1 globally for all trajectories. Conversely, increasing $Y_e$ by 0.1 reduces the lanthanide fractions by a factor of 4, but at the same time reduces the heating rate at all times by about $20-40\,\%$. Hence, the emission is only marginally more luminous, but its duration is $\sim 20\,\%$ shorter and the spectrum shifted towards the blue end of the near-IR band. The fact that the impact of these $Y_e$ modifications is of comparable magnitude as the variations due to different EoSs or binary mass ratios implies that neutrino interactions must be properly accounted for in models that shall be used to reliably infer the ejecta properties.

\item None of our investigated models produces a transient compatible with the blue or red component of AT2017gfo that accompanied GW170817. Agreement with the high spectral temperatures and high luminosities of the early, blue component of AT2017gfo would require significantly smaller lanthanide mass fractions, and therefore opacities, than found for our models. While the low spectral temperatures of our models are roughly compatible with the late, red component of AT2017gfo, the short duration and low luminosity of our kilonova signals would call for a significantly higher ejecta mass than the $0.009\,M_\odot$ of our most massive case. However, our results are not conclusive regarding the role of dynamical ejecta in AT2017gfo as they cover only a small set of EoSs and binary parameters allowed by GW170817 and do not include secular post-merger ejecta.

\item We found that one-zone kilonova models, which assume that the nucleosynthesis results for a single trajectory are representative for the bulk of the ejecta, tend to systematically underestimate the lanthanide fractions, and therefore the opacities, compared to models adopting an ensemble of trajectories that samples a range of thermodynamic conditions. Moreover, the heating rates predicted by one-zone models may depend rather sensitively on the particular choice of thermodynamic conditions along the trajectory and on the employed nuclear-physics input, in contrast to which the average heating rate for an ensemble of trajectories was found to be more robust with respect to the variation of global parameters.
\end{itemize}

The fact that our kilonova models based on single-component dynamical-ejecta configurations are unable to reproduce the blue or the red component of AT2017gfo is in line with previous analyses, such as \citet[][]{Siegel2019r, Nedora2021p}. We note, however, that these works are based on a comparison with the Arnett-type (i.e. essentially spherically symmetric one-zone) model of AT2017gfo by \citet{Villar2017a}. Since radiative transfer models can predict quite different peak properties for specific observer directions than one-zone models -- particularly in the case of multi-component ejecta \citep{Korobkin2021e} -- the masses, bulk velocities, and bulk opacities found in \citet{Villar2017a} for the different components of AT2017gfo should be taken with care.

The role of the dynamical ejecta in AT2017gfo may not be constrained to the minor effects suggested by the single-component models considered in our study. First, our set of models considers only very few EoSs and binary mass ratios, while more extreme configurations may lead to significantly more massive and/or lanthanide-poor ejecta and correspondingly brighter and blue-shifted kilonova signals. Another, more general argument may be connected to the tendency\footnote{We note that kilonova studies \citep[e.g.][]{Wollaeger2018a} informed by Newtonian merger models \citep[e.g.][]{Rosswog2014} often consider purely toroidal geometries to represent the dynamical ejecta. This is because matter ejection into polar directions as a result of collision shocks and corresponding heating is typically less pronounced in Newtonian simulations compared to the general relativistic case, mainly because of effectively weaker effects by the gravitational interaction of the colliding bodies.} that dynamical ejecta are expelled in all directions, albeit with a greater mass loading in equatorial than polar directions \citep{Bauswein2013, Sekiguchi2015a, Radice2018b, Kullmann2021x}. The dynamical ejecta should therefore overcast all non-relativistic post-merger ejecta to a substantial extent, such that a reprocessing of the emission from interior ejecta by the enveloping dynamical ejecta can be naturally expected, in particular at early times ($t\la 1-1.5\,$d) when most of the dynamical ejecta are still optically thick. The model by \citet{Kawaguchi2018a}, which is based on an analytic distribution of multi-component ejecta, supports this notion, and it demonstrated that the blue component of AT2017gfo can be produced even by a lanthanide-rich dynamical-ejecta component, if the latter is subject to heating by photons coming from an enshrouded cloud of lanthanide-poor post-merger ejecta. Even though this model did not adopt a self-consistent ejecta configuration from a hydrodynamic simulation, the results highlight the difficulty of connecting specific kilonova (e.g. blue or red) components to individual (e.g. dynamical or viscous or neutrino-driven) ejecta components, because a kilonova from multiple ejecta components may not just be the sum of kilonovae from single-component ejecta models.

Keeping the aforementioned aspects in mind, we also mention some alternative scenarios that could explain the blue component of AT2017gfo. The partial disruption of the dynamical ejecta due to a relativistic jet and its corresponding energy injection have been considered as a possible way to make the ejecta shine brighter and at higher temperatures \citep{Nativi2020a, Klion2021d}. Moreover, neutrino oscillations could change the nucleosynthesis signature of the ejecta and reduce their lanthanide content \citep[e.g.][]{Zhu2016r, George2020a}. Apart from the dynamical ejecta and subsequent neutrino-driven winds from the surface of the hyper-massive NS, additional fast and lanthanide-poor material could be ejected in connection to magnetic-field effects \citep[e.g.][]{Metzger2018b, Mosta2020c, Shibata2021c} or in the form of a spiral-wave wind powered by an $m=1$ mode \citep{Nedora2019y}. Future, more refined neutrino-magneto-hydrodynamic models of the merger and its remnant as well as self-consistent kilonova calculations based on these models will have to reduce the large uncertainties that, as of yet, prohibit a conclusive assessment of the constituents that shaped the kilonova in GW170817.

Although the number of investigated models is small, our results provide a rough idea about the level of variation that is connected to different possible EoSs, binary mass ratios, and uncertainties in the treatment of neutrino transport. In our models the composition, and therefore the opacity and heating rate, depend only mildly on the EoS and mass ratio. Partially as a result of this circumstance, the analytic scaling formulae, Eqs.~(\ref{eq:tpeakscaling}),~(\ref{eq:Lpeakscaling}),~and~(\ref{eq:Tpeakscaling}), describe relative differences of the peak properties between models with different ejecta masses, $\Mej$, and different bulk velocities, $\langle v\rangle$, reasonably well. Hence, the dependence of the kilonova peak on global parameters of the merger can, to first order, be obtained by feeding the results for $\Mej$ and $\langle v\rangle$ known from detailed merger simulations \citep[e.g.][]{Bauswein2013, Radice2018b, Nedora2020r, Kruger2020a} into Eqs.~(\ref{eq:tpeakscaling}),~(\ref{eq:Lpeakscaling}),~and~(\ref{eq:Tpeakscaling}). In a similar manner, the analytic formulae can also be used to approximately translate yet existing uncertainties from neutrino-hydrodynamic and nucleosynthesis modeling into error bars for the kilonova signal. We refer to Sect.~5 of Part I for a discussion of uncertainties and detailed comparison with literature results for the ejecta properties.

Finally, we summarize the shortcomings of our study that should be improved upon in future work: Since the M1 approximation is less reliable in the optically thin regime, our models do not have the same predictive power as kilonova calculations based on Boltzmann solvers, particularly concerning the emission into specific directions. Also, we only include the early, dynamical ejecta here, but additional ejecta components produced by the merger remnant should be included as well, ideally accounting for the dynamical and radiative interactions between components. Moreover, the transition between the phase of matter ejection and homologous expansion is not included but should be accounted for to ensure a correct mass distribution as input for the kilonova calculation. Finally, our phenomenological treatment of the opacities is not based on atomic properties, which themselves depend on the local thermodynamic conditions. Our scheme could readily be improved by using an opacity framework based on atomic-physics models.

\section*{Acknowledgments}
OJ, AB, and CC acknowledge support by the European Research Council (ERC) under the European Union's Horizon 2020 research and innovation programme under grant agreement No. 759253 and by the - Project-ID 279384907 - Sonderforschungsbereich SFB 1245 by the Deutsche Forschungsgemeinschaft (DFG, German Research Foundation). OJ was partially supported by JSPS Grants-in-Aid for Scientific Research KAKENHI (A) 19H00693 and by the Interdisciplinary Theoretical and Mathematical Sciences Program (iTHEMS) of RIKEN. SG acknowledges financial support from F.R.S.-FNRS (Belgium). This work has been supported by the Fonds de la Recherche Scientifique (FNRS, Belgium) and the Research Foundation Flanders (FWO, Belgium) under the EoS Project nr O022818F. AB was additionally supported by - Project-ID 138713538 - SFB 881 (``The Milky Way System'', subproject A10) and by the State of Hesse within the Cluster Project ELEMENTS. At Garching, funding by the European Research Council through Grant ERC-AdG No.~341157-COCO2CASA and by the DFG through SFB-1258 ``Neutrinos and Dark Matter in Astro- and Particle Physics (NDM)'' and under Germany's Excellence Strategy through Cluster of Excellence ORIGINS (EXC-2094)---390783311 is acknowledged. OJ acknowledges computational support by the HOKUSAI supercomputer at RIKEN/Japan and by the Max-Planck Computing and Data Facility (MPCDF) at Garching/Germany. The nucleosynthesis calculations benefited from computational resources made available on the Tier-1 supercomputer of the Fédération Wallonie-Bruxelles, infrastructure funded by the Walloon Region under the grant agreement n$^\circ$1117545 and the Consortium des Équipements de Calcul Intensif (CÉCI), funded by the Fonds de la Recherche Scientifique de Belgique (F.R.S.-FNRS) under Grant No. 2.5020.11 and by the Walloon Region. 

\emph{Data availability:} The data underlying this article will be shared on reasonable request to the corresponding author.



\appendix

\section{Opacity calibration}\label{sec:opacity-calibration}

\begin{figure*}
\includegraphics[width=0.49\textwidth]{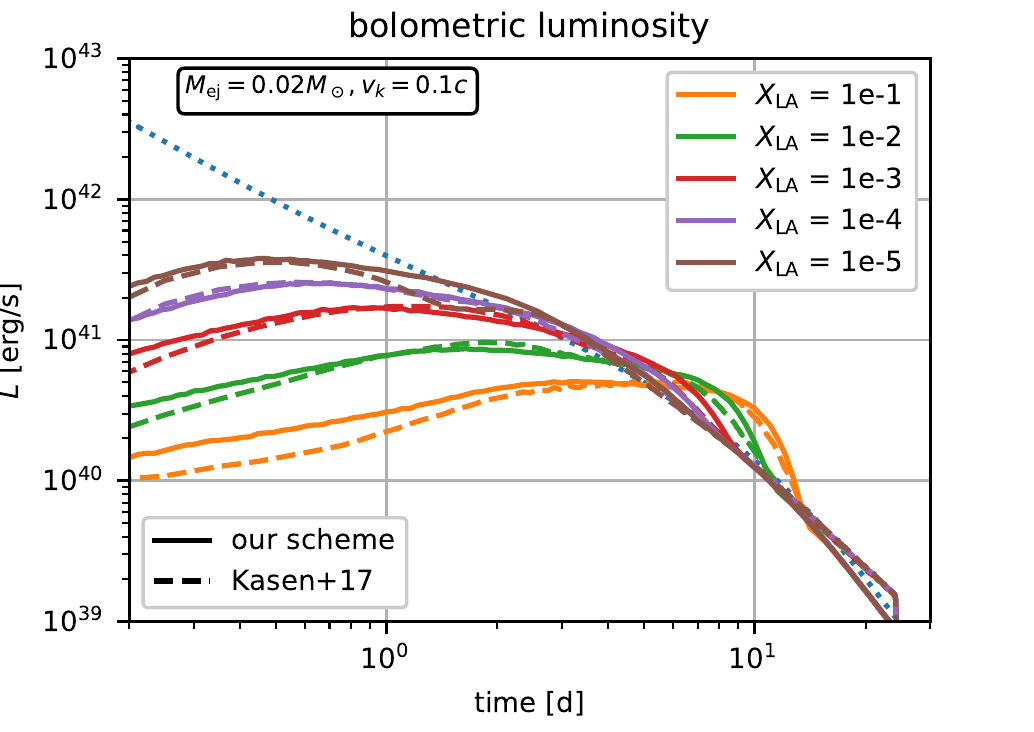}
\includegraphics[width=0.49\textwidth]{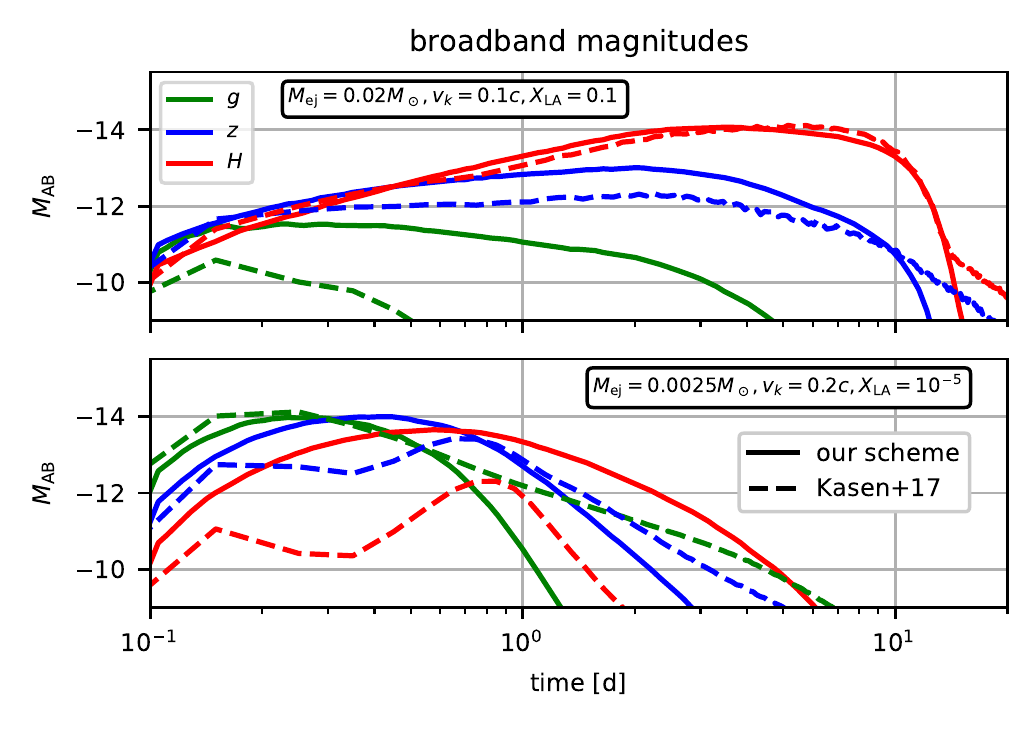}
\caption{\emph{Left panel:} Results of our opacity calibration using a set of bolometric light curves from \citet{Kasen2017a}. Solid lines refer to results obtained with our scheme, while dashed lines are the reference results from \citet{Kasen2017a}. Model parameters are listed in the insets. The blue dotted line denotes the effective heating rate after thermalization. \emph{Right panel:} Comparison of broadband magnitudes in three exemplary frequency bands for two qualitatively different models. The top panel shows results for a model leading to a near-IR kilonova, while the model in the bottom panel produces a kilonova peaking rather in the optical regime. Despite quantitative differences, our scheme reproduces the peaks in the correct bands.}
\label{fig_kasenlum}
\end{figure*}

Our kilonova treatment avoids the complexity connected to a microphysically consistent opacity, $\kappa$, which would need to account for thousands of transition lines between energy levels -- most of them poorly known so far -- of many isotopes in dependence of the detailed composition, density, temperature, and degree of ionization. Instead, we employ a simplified, grey version that parametrizes the opacity in terms of the lanthanide mass fraction, $\XLA$, and temperature, $T$; see Eqs.~(\ref{eq:kappatot}-\ref{eq:kappatem}). The form of $\kappa$ was calibrated\footnote{In order to prevent confusion, we note that a similar calibration was already presented in the Appendix of \citet{Just2021a}, which suggested a slightly different form of $\kappa(\XLA,T)$. The reason is that \citet{Just2021a} did not correct their luminosities for frame-dependent effects as we do here (cf. Sect.~\ref{sec:extr-observ-light}) and, hence, needed different opacities in order to reproduce the same reference results. We recommend using the newer version presented here.} in a way to reproduce, as accurately as possible, a set of bolometric light curves by \citet{Kasen2017a} using a state-of-the-art opacity treatment, which differ by the contained mass of lanthanides. In Fig.~\ref{fig_kasenlum}, left panel, we provide the light curves based on which we calibrated. The ejecta mass and bulk velocity for this reference case are $0.02\,M_\odot$ and $0.1\,c$, respectively, while we refer to \citet{Kasen2017a} for the precise form of the analytic mass distribution. For this test we use a radioactive heating rate that is presumably similar, though possibly not exactly identical, to the one used in \citet{Kasen2017a}. It is taken from \citet{Lippuner2015a} and given by $q_{\mathrm{rad}}$\,[erg\,g$^{-1}$\,s$^{-1}$]$=(1.0763\times 10^{10} t_{\mathrm{day}}^{-1.518}+9.5483\times 10^9 e^{-t_{\mathrm{day}}/4.947})$ with $t_{\mathrm{day}}$ being the time in units of days. The homogeneous thermalization efficiency is computed, as in \citet{Kasen2017a}, using the analytic fit formula as function of the ejecta mass and velocity provided by \citet{Barnes2016a}.

A comparison of the peak behavior of broadband light curves provides another consistency check for our treatment. In Fig.~\ref{fig_kasenlum} we compare the broadband light curves in the $g$, $z$, and $H$ bands for two models. The model in the top panel of Fig.~\ref{fig_kasenlum} assumes rather massive, relatively slow, and lanthanide-rich ejecta, and the peak is in the near-IR ($H$) band. The model in the bottom panel assumes less massive ejecta with higher velocities and a low lanthanide fraction, and the peak is therefore at higher spectral temperatures, namely in the $g$ and $z$ bands. Albeit quantitative differences are noticeable, our scheme can reproduce the basic peak behavior quite well. This result lends credibility to the broadband behavior obtained in the dynamical ejecta models investigated in this paper.

\section{Sensitivity to initial conditions}\label{sec:sens-init-cond}

\begin{figure}
\includegraphics[width=0.49\textwidth]{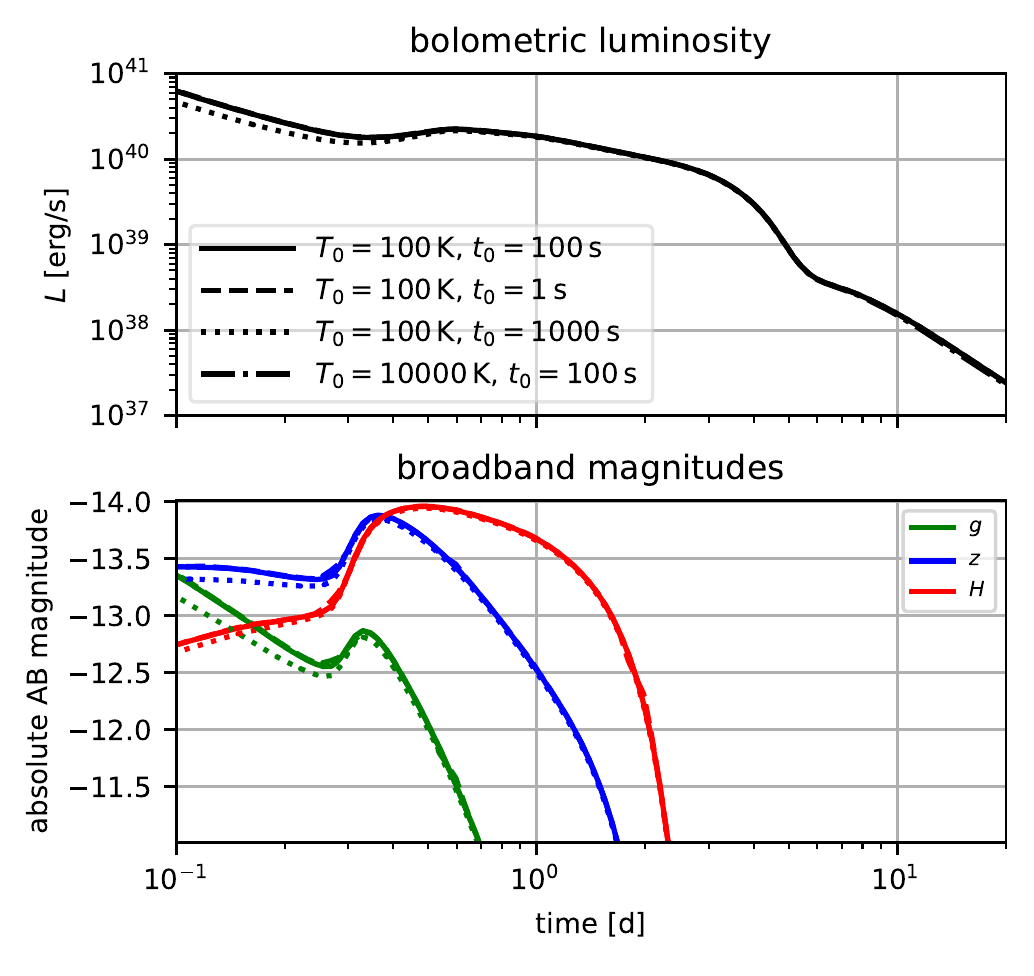}
\caption{Comparison of bolometric luminosities \emph{(top panel)} and broadband magnitudes \emph{(bottom panel)} for kilonova calculations of model DD2-135135 initialized with different gas temperatues, $T_0$, and at different times, $t_0$. Solid lines correspond to the standard case used for all models in the main part of this study. The lines corresponding to $t_0=100\,$s and $T_0=10000\,$K lie on top of the solid lines. Small differences at early times are only visible for the model initialized at $t_0=1000\,$s.}
\label{fig_initcon}
\end{figure}

As argued in Sect.~\ref{sec:lightc-calc}, the light curve at $t\ga 0.1\,$d is fairly independent of the choice of the initial conditions -- namely the initialization time, $t_0$, as well as the initial distributions of radiation energy and gas temperature -- as long as $t_0$ is chosen sufficiently early for the system to relax well before $t\sim 0.1\,$d into the correct state of balance between radioactive heating and adiabatic expansion. In this Appendix we back this argument by showing in Fig.~\ref{fig_initcon} light curves for variations of the fiducial model, DD2-135135, using different initial conditions. All models in the main part of this study are initialized at $t_0=100\,$s with a constant temperature of $T_0=100\,$K. The excellent agreement of the fiducial model with one model using instead $t_0=1\,$s and with another model using $T_0=10000\,$K supports the robustness of our results with respect to the initial conditions. Figure~\ref{fig_initcon} also reveals that a later initialization time of $t_0=1000\,$s would have led to a slight underestimation of the luminosities, though only at early times $t\la 0.5\,$d.

\end{document}